\documentclass[%
 reprint, 10pt,
nofootinbib,
 amsmath,amssymb,
 aps
]{revtex4-1}

\pdfoutput=1
\usepackage{graphicx}
\usepackage[utf8]{inputenc}
\usepackage{dcolumn}
\usepackage{bm}

\usepackage[a4paper,top=2cm,bottom=2cm,left=3cm,right=2cm]{geometry}
\usepackage{amssymb,graphicx}
\usepackage{epstopdf}
\usepackage{amsmath,amsfonts}
\usepackage{mathtools}
\usepackage{hyperref}
\usepackage[dvipsnames]{xcolor}
\usepackage{caption}
\usepackage{subcaption}
\usepackage{verbatim}
\usepackage[utf8]{inputenc}

\usepackage[T1]{fontenc}

\def\e{{\rm e}}

\newcommand{\be}{\begin{equation}}
\newcommand{\ee}{\end{equation}}
\newcommand{\bea}{\begin{eqnarray}}
\newcommand{\eea}{\end{eqnarray}}
\newcommand{\bg}{\begin{gather}}
\newcommand{\eg}{\end{gather}}
\newcommand{\bseq}{\begin{subequations}}
\newcommand{\eseq}{\end{subequations}}

\renewcommand{\ln}{\mathop{\rm ln}\nolimits}

\newcommand{\M}{M_{\rm Pl}}

\renewcommand{\L}{\mathcal L}

\begin{document}

\preprint{Imperial/TP/2019/LA/01}Imperial/TP/2019/LA/01

\bigskip

\title{\large 
Equivalence Principle on Cosmological Backgrounds\\
 in Scalar-Tensor Theories%
%
%
%
%
}

\author{Lasma Alberte}%
 \email{lalberte@ic.ac.uk}
\affiliation{Theoretical Physics, Blackett Laboratory, Imperial College, London, SW7 2AZ, U.K. }%

\begin{abstract}

We study the extent up to which the equivalence principle is obeyed in models of modified gravity and dark energy involving a single scalar degree of freedom. We focus on the effective field theories of dark energy describing the late time acceleration in the presence of ordinary matter species. In their covariant form these coincide with the Horndeski theories on a cosmological background with a slowly varying Hubble rate and a time-dependent scalar field. 
We show that in the case of an exactly de Sitter universe both the weak and strong equivalence principles hold. This occurs due to the combination of the shift symmetry of the scalar and the time translational invariance of de Sitter space. 
%
When generalized to Friedmann--Robertson--Walker cosmologies (FRW) we show that the weak equivalence principle is obeyed for test particles and extended objects in the quasi-static subhorizon limit. We do this by studying the field created by an extended object moving in an FRW background as well as its equation of motion in an external gravitational field. We also estimate the corrections to the geodesic equation of the extended object due to the approximations made. 

\end{abstract}

\maketitle

\section{Introduction and motivation}
Over the past two decades it has become clear that our universe today is in a phase of accelerated expansion driven by an unknown component---the dark energy. Although the cosmological constant provides an almost perfect fit to the observational data and is theoretically the most economic solution to this problem, its unnaturally small value has forced us in the quest of finding other, more satisfying dark energy models. A large fraction of the recent proposals rely on modifying gravity at large distances and fall in the class of the effective field theories (EFTs) of dark energy~\cite{Gubitosi:2012hu}. This framework unifies competing models of modified gravity that in a given window of energy scales involve only one additional scalar degree of freedom with respect to General Relativity (GR). Written directly for the cosmological perturbations, the EFT of dark energy can be interpreted as a low energy effective theory for the Goldstone boson of broken time translations. In the spirit it is very similar to the EFT of inflation where the scalar degree of freedom is usually the inflaton field breaking the de Sitter invariance of the early universe \cite{Cheung:2007st}. This approach captures most of the various phenomenological effects arising in the different dark energy models and allows to discriminate between them. 

A typical feature of the dark energy models involving a light scalar degree of freedom is to introduce order one modifications of gravity on large scales. While it is an acceptable and seeked for effect on cosmological scales, these theories are highly constrained by short distance observations. In particular, any such modified gravity theory needs to have a working screening mechanism, such as the Vainshtein mechanism \cite{Vainshtein:1972sx,Babichev:2013usa}, in order to recover the standard GR results at short scales. An important subclass of the EFTs of dark energy implementing the Vainshtein mechanism is the Horndeski theories, originally formulated in \cite{Horndeski:1974wa} and later rediscovered in \cite{Deffayet:2009wt,Deffayet:2011gz}. These are the most general scalar--tensor theories with second order metric and scalar field equations of motion and have been extensively studied in the recent years.

A novel avenue for testing the modified gravity models has opened up due to the recent observation of the gravitational wave emission during the black hole and neutron star merger events \cite{Abbott:2016blz,TheLIGOScientific:2017qsa,Monitor:2017mdv}. The significance of these observations lies in the fact that it offers the possibility to test gravity in the previously inaccessible highly relativistic, strong-field regime. For instance, the observation of gravitational waves from the binary neutron star merger, arriving simultaneously with their optical counterpart, allowed to constrain the speed of gravitational waves to coincide with the speed of light up to deviations of order $10^{-15}$~\cite{TheLIGOScientific:2017qsa,Monitor:2017mdv}. This observation alone ruled out a wide range of operators within the EFTs for dark energy that were predicting subluminal propagation speeds for the gravity waves \cite{Creminelli:2017sry, Sakstein:2017xjx,Baker:2017hug,Ezquiaga:2017ekz}. A possible caveat of this conclusion might lie in the fact that the energy scale at which the gravitational waves were observed lies very close to the strong coupling scale of many of the dark energy models in question. Being only effective field theories they would naturally receive corrections above the strong coupling scale that could render the speed of propagation of the gravity waves to lie close to the speed of light \cite{deRham:2018red}. We shall comment more on this point below. 

Another possible way of testing GR with the help of the gravitational wave observations is by studying the strong-field effects. These include, in particular, the quasi-normal mode spectrum \cite{Dong:2017toi,Tattersall:2018nve,Franciolini:2018uyq,Witek:2018dmd,Mirbabayi:2018mdm} and the violations of the strong equivalence principle leading to the emission of the dipolar radiation \cite{Berti:2015itd}. 

Motivated by these advances in observational cosmology, in this work we shall focus our attention on the possible violations of the equivalence principle within the modified theories of gravity. Equivalence principle in one of its simplest forms can be stated as the equality between the inertial and gravitational masses of a given object \cite{Will:2005va}. When referring to objects with small self-gravitational energy one talks about the \emph{weak} equivalence principle. In turn, the \emph{strong} equivalence principle requires it to hold also for objects with a large fraction of their mass consisting of the gravitational binding energy, like black holes and neutron stars. 

The violations of the equivalence principle in modified gravity theories arise due to the coupling of the new gravitational degrees of freedom---a single scalar field in our case---to  matter. It is in general believed that in theories where the extra fields are only coupled to the metric and bear no direct coupling to the matter fields the weak equivalence principle is obeyed (for exceptions, see \cite{Hui:2009kc}). However, for gravitational bodies with strong self-gravity the additional fields due to the modification of gravity will still effectively couple to matter \cite{Will:2005va}. Even though at leading order the coupling to matter is absent, at higher orders in perturbations it will arise because of the non-minimal coupling to the metric, which in turn is coupled to matter. As a result the body's gravitational mass will depend on the additional fields. This leads to the fact that strongly gravitating bodies follow trajectories that depend on their internal structure and composition, known as the Nordtvedt effect \cite{Nordtvedt:1968qr}. 

The physical consequence of the Nordtvedt effect is the appearance of the fifth forces and this has profound implications on the gravitational wave emission. In particular, these fifth forces increase the total power of the gravitational wave emission as a result of which the orbits of inspiraling binaries shrink faster than in GR. One can show that this happens due to the dipole emission of the gravitational radiation (see \cite{Barausse:2017gip} and references therein). The monopole and dipole emission are therefore known as the red flags for scalar-tensor theories. In fact, even before the observation of the gravitational waves, the previous data on the observed mildly relativistic binary pulsars was providing the most stringent tests~\cite{Freire:2012mg} on the `classic' scalar-tensor theories of the Brans--Dicke type~\cite{Brans:1961sx}. 

The recent observations by LIGO and Virgo will provide an unseen level of precision for constraining the violations of the equivalence principle. It is therefore timely to review our understanding of the extent up to which such violations occur in modified gravity theories. The focus of this work will be to investigate the effects due to the cosmological expansion---an issue that up to the best of our knowledge has not been explored in this context in the previous literature. 

 Let us also emphasize that we will not consider explicit violations of the EP. Such violations usually arise in theories where different matter fields are coupled to gravity with different strength already at the level of action. Instead we shall assume a universal coupling of our scalar-tensor theory to all matter fields. 

There are a several earlier works on the equivalence principle in scalar-tensor theories that need to be mentioned before we present our results. 

The first class of works is related to the no-hair theorem for black holes in scalar-tensor theories. It is known that in GR black holes obey a no-hair theorem that can be extended also to the simplest of scalar-tensor theories with canonical derivative interactions for the scalar field \cite{Bekenstein:1971hc,Bekenstein:1995un}.  In \cite{Hui:2012qt}, the proof of the no-hair theorem was further extended to the case of Galileons on asymptotically flat backgrounds \cite{Nicolis:2008in}. The main assumptions needed for the proof were the shift symmetry of the scalar field action, as well as the spherical symmetry and time independence of the black hole and scalar field solution. As a consequence of the theorem the black hole does not admit any non-trivial profile of the Galileon field. On the other hand ordinary matter does source a scalar field profile and thus experiences the fifth force. Hence, a test particle and a black hole would move on different geodesics in any scalar-tensor theory that can be reduced to Galileons in some relevant limit. This suggests the violation of the strong equivalence principle in these theories. It was later shown in \cite{Sotiriou:2013qea,Sotiriou:2014pfa} that the no-hair theorem of \cite{Hui:2012qt} is violated in a more general set-up by also including a coupling of the scalar field to the Gauss-Bonnet term. Such a coupling appears generically in Horndeski theories \cite{Horndeski:1974wa,Deffayet:2009wt,Deffayet:2011gz}. 

The situation turns out to be different for static spherically symmetric and asymptotically flat \emph{star} solutions (\emph{i.e.} solutions that are regular everywhere in the spacetime and possess no horizon).  In \cite{Lehebel:2017fag,Barausse:2017gip} it was shown that stars in shift-symmetric Horndeski theories obey a no-hair theorem and thus have a zero scalar charge. A similar conclusion was made earlier in \cite{Barausse:2015wia} where it was found that there is no emission of dipolar radiation in such theories. Let us emphasize that all the works mentioned in the context of no-hair theorems work on asymptotically flat backgrounds and make use of \emph{static} scalar field profiles. Relaxing one of these assumptions allows one to evade the no-hair theorems of \cite{Hui:2012qt,Lehebel:2017fag}. For instance, allowing for (Anti) de Sitter asymptotics can lead to hairy black hole and star solutions \cite{Rinaldi:2012vy,Minamitsuji:2013ura} in the Fab Four \cite{Charmousis:2011bf} subclass of the Horndeski theories. A review on asymptotically flat black hole solutions with non-trivial scalar hair can be found in \cite{Herdeiro:2015waa}.

A line of research closely related to the no-hair theorems is the existence of static spherically symmetric solutions in scalar-tensor theories. There is a lot of literature on such solutions by using both time independent and time dependent scalar field profiles leading to both asymptotically flat and asymptotically (Anti-)de Sitter solutions. In this work we shall focus on static spherically symmetric field configurations on top of cosmological backgrounds and \emph{time dependent} scalar field profiles. In the context of black hole solutions in scalar-tensor theories these were first used in \cite{Babichev:2012re,Babichev:2013cya} giving both asymptotically flat and asymptotically de Sitter solutions in the vacuum. Phenomenologically viable asymptotically flat neutron star solutions in the presence of matter were found in \cite{Cisterna:2015yla}. Of particular relevance for this paper is the work by Babichev et al. \cite{Babichev:2016fbg} where static spherically symmetric asymptotically de Sitter solutions in the cubic Galileon theory were studied. We shall discuss these in more detail in Section~\ref{sec:equiv_dS}. 

The second class of works related to the equivalence principle  in modified gravity theories is concerned with objects with negligible gravitational self-energy \cite{Hui:2009kc,Hui:2010dn}. These are only dealing with the question of possible violations of the \emph{weak} equivalence principle. In particular, it was shown in \cite{Hui:2009kc} that in scalar-tensor theories with Chameleon screening mechanism \cite{Khoury:2003aq,Khoury:2003rn}, such as the Brans--Dicke theory \cite{Brans:1961sx}, there are $\mathcal O(1)$ violations to the weak equivalence principle. On the other hand, it was argued that in theories with Vainshtein screening mechanisms such violations are absent. These results were obtained by studying the motion of an extended object in an external gravitational field by the method of Einstein, Infeld, Hofman \cite{Einstein:1938yz,damour}. We shall review this approach in great detail in Section~\ref{sec:geodesics}. Finally, let us also mention that in \cite{deRham:2012fw} the Vainshtein mechanism in conformally coupled cubic Galileon theory was studied for binary systems. It was found that both the monopole and dipole radiation vanishes at the leading order due to successful Vainshtein screening, thus confirming that the weak equivalence principle is only violated once the relativistic corrections are included.

The paper is organized as follows. In Section~\ref{sec:general} we shall present the effective field theory action for dark energy that we shall be using in the rest of this work and derive the equations of motion for the relevant degrees of freedom. We shall also introduce the various types of objects studied: test particles, extended objects and black holes. In Section~\ref{sec:shift} we shall then discuss the extent up to which the shift symmetry is preserved in the scalar--tensor theories under question. We shall review the no-hair theorem for the black holes in asymptotically flat Galileon theories relying on the conservation of the shift symmetry current. In~\ref{sec:currentfrw} we shall then show how the argument is modified on FRW backgrounds and argue that the weak equivalence principle holds for screened extended objects in Sec.~\ref{extended}. 

An argument that both the strong and the weak equivalence principles hold in the special case of de Sitter universe is presented in Section~\ref{sec:equiv_dS}. By an appropriate choice of the coordinate system, we derive the solutions for the asymptotic field profiles for the fields far away from the object in Sec.~\ref{sec:solution}. In particular, we show that the scalar field profile is non-trivial and only depends on the ADM mass of the object, thus corresponding to a secondary hair. We present the explicit asymptotic solution and show that it approaches the Schwarzschild-de Sitter solution in the subhorizon limit. In Sec.~\ref{sec:generalize} we present a framework for studying the violations of the strong equivalence principle due to FRW-type of deviations from de Sitter spacetime. 

Finally, in Section~\ref{sec:geodesics} we study the motion of extended objects on FRW background by the method of Einstein, Infeld and Hoffmann \cite{Einstein:1938yz,damour}. As in Sec.~\ref{extended} we arrive at the conclusion that the weak equivalence principle holds for extended objects. We compute the violations due to the departures from the full Vainshtein screening regime in Sec.~\ref{sec:violations}. We conclude in Sec.~\ref{conclusions}.

\section{Generalities}\label{sec:general}
\subsection{The action}
We shall consider an effective field theory for cosmological perturbations around an FRW background driven by a time-dependent scalar field. The constant time hypersurfaces are parametrized by the scalar field and in the so-called unitary gauge the time and the scalar coincide. Around the metric $ds^2= -dt^2+a(t)^2d\vec x^2$ the EFT action written in this gauge reads \cite{Gubitosi:2012hu,Gleyzes:2014dya}:
\be
\begin{split}
\label{act_unit}
S&=  \int d^4x\,\sqrt{-g}\left[\frac{\M^2}{2}f(t)R-\Lambda(t)-c(t)g^{00}\right.\\
&\left.+\frac{m_2^4(t)}{2}\left(\delta g^{00}\right)^2-\frac{m_3^3(t)}{2}\delta K\delta g^{00}\right
]\\&+\int d^4x\,\sqrt{-g}\,\mathcal L_{\text m}(\psi_{\text m},g_{\mu\nu})\,.
\end{split}
\ee
Here $\delta g^{00}=g^{00}+1$, $\delta K=K-3H$, $H=\dot a/a$ is the Hubble parameter, and $K=\nabla_\alpha n^{\alpha}$ is the extrinsic curvature of surfaces with $t=\text{const}$. The unit vector $n_\mu$ is perpendicular to the constant time slicing, parametrized through the scalar field as $n_\mu\equiv -\partial_\mu\phi/\sqrt{-(\partial\phi)^2}$. The above action does not contain an explicit dependence on the scalar field perturbations $\pi$ since these are set to zero in the unitary gauge where $\phi=\M^2t$. The function $\mathcal L_{\text m}$ stands for the Lagrangian density for some matter fields collectively denoted by $\psi_{\text m}$, that are assumed to only be coupled to the metric $g_{\mu \nu}$, and not to the scalar field. In other words, the action \eqref{act_unit} is written in the Jordan frame. 

The action \eqref{act_unit} contains contributions to all orders in perturbations. However, only the first three terms contribute to the background equations. The first non-zero contribution from the $m_2^4,m_3^3$ terms on the second line appears at quadratic order. 
In fact, the action~\eqref{act_unit} is not the most general quadratic low-energy effective action preserving the spatial diffeomorphisms, and a couple of additional terms should be included \cite{Gleyzes:2014dya}:
\begin{equation}\label{m4}
+m_4^2(t)(\delta K^\mu_\nu\delta K^\nu_\mu-\delta K^2)+\frac{\tilde m_4^2(t)}{2}R\,\delta g^{00}\,.
\end{equation}
However, these terms have lately been in an increasing tension with observations. In particular, the observation of the gravitational waves from the neutron star--black hole binary merger has constrained  the speed of gravitational waves to coincide with the speed of light \cite{TheLIGOScientific:2017qsa,Monitor:2017mdv}.  This observation then enforces the parameter choice $m_4=0$ \cite{Creminelli:2017sry,Sakstein:2017xjx,Baker:2017hug,Ezquiaga:2017ekz}. In Horndeski theories, $m_4=\tilde m_4$, and thus both of the terms above are excluded. In principle, the operator $R\,\delta g^{00}$ is still allowed in beyond Horndeski theories where $m_4\neq\tilde m_4$ \cite{Gleyzes:2014dya}. However, it was pointed out in \cite{Creminelli:2018xsv} that the presence of these operators would result in a graviton decay into the scalar field particles. The fact that we do observe the gravitational waves then puts a bound on the decay rate, setting $\tilde m_4=0$. It should, however, be kept in mind that the frequency of that particular neutron star merger event lies very close to the strong coupling scale of the dark energy models \eqref{act_unit}, \eqref{m4}. This means that, in principle, new operators, coming from a hypothetic UV completion of these theories can affect the speed of gravity waves and the decay rate on that scale  \cite{deRham:2018red}. Hence, ruling out the operators \eqref{m4} might be somewhat premature; we will however stick to the more conservative attitude and not consider them in this work.
 
We also note that the function $f(t)$ typically arises in Brans-Dicke type theories \cite{Brans:1961sx}. Since these are known to introduce $\mathcal O(1)$ violations to both the strong \cite{Hawking:1972qk} and the weak equivalence principle \cite{Hui:2009kc}, we shall not consider it in the following and set $f(t)\equiv 1$. 

In the remainder of this work we shall be working with the covariant form of the unitary gauge action~\eqref{act_unit}. Its diffeomorphism invariance can be restored by the usual St\"uckelberg trick: we perform an infinitesimal time diffeomorphism $t\to t+\pi(t,\vec x)$ and introduce a scalar field
\be\label{phi}
\frac{\phi}{\M^2}=t+\pi(t,\vec x)\,.
\ee
It is important to emphasize the fact that it is possible to restore the full diffeomorphism invariance by introducing a single scalar field. This shows that the low-energy dynamics of the perturbative action \eqref{act_unit} is entirely determined by one single additional degree of freedom compared to pure GR. This constrains our analysis to this type of models. While it includes, for instance, the Horndeski theories, it does not include models that contain more than one degree of freedom in the IR. A typical example of what these theories do not describe would be massive gravity where~$\pi$ arises as the helicity-0 mode of a massive spin-2 multiplet \cite{ArkaniHamed:2002sp,deRham:2010ik,deRham:2010kj}. This was in fact clear already from the action \eqref{act_unit} since by construction it is invariant under the spatial diffeomorphisms. In contrast, the massive gravity action, even in the decoupling limit where it is dominated by the scalar-tensor interactions, is known to break the spatial diffeomorphisms. 

Following \cite{Gubitosi:2012hu} and performing the replacements \eqref{replace1}, \eqref{replace2} we find the covariant action to be:
\begin{align}\label{act_cov}
S&=\int d^4x\,\sqrt{-g}\left[\frac{\M^2}{2}R-\Lambda-c\,\frac{(\partial\phi)^2}{\M^4}\right.\nonumber\\
&+\frac{m_2^4}{2}\left(1+\frac{(\partial\phi)^2}{\M^4}\right)^2\\
&-\left.\frac{m_3^3}{2}\left(1+\frac{(\partial\phi)^2}{\M^4}\right)\left(-3H-\frac{3}{2\M^2}\Box\phi\right)\right.\nonumber\\
&+\left.\frac{m_3^3}{4\M^6}\left(1+\frac{(\partial\phi)^2}{\M^4}\right)\left((\partial\phi)^2\Box\phi+\partial^\mu\phi\,\partial_\mu(\partial\phi)^2\right)\right]\nonumber\\
&+\int d^4x\,\sqrt{-g}\,\mathcal L_{\text m}(\psi_{\text m},g_{\mu\nu})\,.\nonumber
\end{align}
When expanded in perturbations it coincides with the unitary gauge action \eqref{act_unit} up to cubic order.
All the quantities $\Lambda, c, m_2, m_3$ that were time dependent in the unitary gauge are now functions of the scalar field, \textit{e.g.} $\Lambda=\Lambda\left(\frac{\phi}{\M^2}\right)$ etc. We have suppressed this in \eqref{act_cov} for brevity.

\subsection{Equations of motion}
In order to assess whether the equivalence principle holds in theories described by the action~\eqref{act_cov}, we shall study the gravitational field \emph{created} by a massive object, as well as its motion in some \emph{external} gravitational field.  We assume that the expansion of the universe is driven by a background dark energy component due to the scalar field and some other perfect fluid-type matter component with $T^{\text m}\,^\mu_\nu=\text{diag}(-\rho_{\text{m}},p_{\text{m}},p_{\text{m}},p_{\text{m}})$. We describe the object on this background by a stress-energy tensor $\delta T^{\text{m}}_{\mu\nu}$. 

We consider metric perturbations around the flat FRW background in the Newtonian gauge:
\be\label{gauge}
ds^2=-e^{2\Phi}dt^2+a^2(t)e^{-2\Psi}\delta_{ij}dx^idx^j\,,
\ee
where the potentials $\Phi$ and $\Psi$ are the only gravitational degrees of freedom. In addition, we have the scalar field perturbation $\pi$. The equations of motion, derived from the action.~\eqref{act_cov}, can be written as:
\begin{align}
&\M^2 G_{\mu \nu}=T^{\phi}_{\mu\nu} +T^{\rm m}_{\mu\nu}\,, \label{Einsteineqs}\\
&\nabla_\mu (T^{\rm m})^\mu_\nu=0\,, \label{matter_cons}\\
&\nabla_\mu \left[\left(T^{\phi}\right)^\mu_\nu-\M^2 G^\mu_\nu\right]=\frac{1}{\sqrt{-g}}\frac{\delta S}{\delta\phi}\partial_\nu\phi\label{eqpi}\,.
\end{align}
The first equation comes from the variation of the total action with respect to the metric $g^{\mu\nu}$. We define the matter stress-energy tensor in the canonical way as $T^{\rm m}_{\mu\nu}\equiv\frac{-2}{\sqrt{-g}}\frac{\delta S_{\text m}}{\delta g^{\mu\nu}}$, and the stress-energy tensor associated to the scalar field, $T^{\phi}_{\mu\nu}$, is implicitly defined through Eq.~\eqref{Einsteineqs}. We give the full covariant expression for the variation of the action~\eqref{act_cov} with respect to the metric in Appendix~\ref{app:emtcov}. The second equation above comes from assuming that the matter action is only coupled to the metric and is invariant under spacetime diffeomorphisms. The last equation is the conservation equation---a consequence of the diffeomorphism invariance of the action. The $\nu=0$ component of it is equivalent to the equation of motion for the scalar field~$\phi$.

The background equations of motion for the FRW metric following from \eqref{Einsteineqs} fix the functions $c$ and $\Lambda$ as
\be
\label{backeqcL}
\begin{split}
c(t)&=-\frac{1}{2}(\rho_m+p_m)-\M^2\dot H\,,\\
\Lambda(t)&=\frac{1}{2}(p_m-\rho_m)+\M^2\left(\dot H+3H^2\right),
\end{split}
\ee
where $\rho_m$ and $p_m$ are the energy density and pressure of the matter component.

For the purpose of this paper it will be convenient to write the Einstein equations for perturbations as:
\begin{align}\label{split}
&\M^2 G^{\rm L}_{\mu\nu}-T^{{\rm L}, \phi}_{\mu \nu}= \tau_{\mu \nu}\\\label{deftau}
& \tau_{\mu \nu}\equiv T^{{\rm NL}, \phi}_{\mu \nu}+\delta T^{\rm m}_{\mu \nu}-\M^2 G^{\rm NL}_{\mu\nu}\,,
\end{align}
where we have introduced $\tau_{\mu\nu}$---an analogue of the Einstein's pseudo stress-energy tensor. The superscript ``L'' means we only keep the linear part of the expressions, while the ``NL'' stands for everything else except for those linear terms. Note that such a definition does not require the validity of a perturbative expansion, which is not guaranteed close to the source. One can nevertheless always define a linear tensor and arbitrarily remove it from the full tensor. This expression assumes that the cosmological background is already solved for, so that no background quantities remain. Let us note that in GR one defines the pseudo stress tensor as $t_{\mu\nu}=\M^2G^{\rm L}_{\mu\nu}=\delta T^{\rm m}_{\mu \nu}-\M^2 G^{\rm NL}_{\mu\nu}$. The virtue of this definition is that due to the linearized Bianchi identities for the Einstein tensor, this stress tensor is conserved in flat space, \emph{i.e.} $\partial_\mu t^\mu\,_\nu=0$. In the case when the gravitational theory is modified by an additional degree of freedom and we are on an FRW background, it is necessary to include also $T^{{\rm L}, \phi}_{\mu \nu}$ in the definition above. Only then $\partial_\mu\tau^\mu\,_\nu=0$ at linear level. We shall discuss this in more detail in Section~\ref{sec:method}.

Given Eq.~\eqref{split}, one can view the nonlinearities of the EFT theory as a source term for the linear solution. The virtue of Eq.~\eqref{split} is then that it is a fully nonlinear equation. Although this might appear as a trivial observation this will, in fact, be important when we will want to distinguish between infinitesimal test particles and objects of finite size and/or large self-gravitational energy. 

To demonstrate this, let us focus on the quasi-static, subhorizon limit (in Fourier space, this corresponds to considering wave-numbers $k\gg aH$ and frequencies $\omega\ll k$). Then, the leading order equations take the form:
\begin{itemize}
\item \textbf{$00$} component of eq.~\eqref{Einsteineqs}
\be
\label{eq00lin}
\frac{1}{a^2}\left[2\M^2\Delta\Psi+m_3^3\Delta\pi\right]=-\tau^0\,_0\,;
\ee
\item \textbf{$ij$} component of eq.~\eqref{Einsteineqs}
\be
\label{eqijlin}
\frac{\partial_i\partial_j-\delta^i_j\Delta}{a^2}\left[\M^2(\Psi-\Phi)\right]=\tau^i\,_j\,;
\ee
\item equation for $\pi$
\be 
\begin{split}
\label{eqpilin}
\frac{\Delta}{a^2} \bigg[m_3^3\Phi+\left(2c+(m_3^3)^{\hbox{$\cdot$}}+Hm_3^3 \right)\pi\bigg]=-\mathcal{E}_{{\rm NL},\pi}\,,
\end{split}
\ee
\end{itemize}
where in the last equation, we have defined $\mathcal{E}_{{\rm NL},\pi}$ as the nonlinear part of the equation of motion for the scalar. The set of equations given above then fully determines the three fields $\Phi,\Psi,\pi$; in particular, their asymptotic behaviour far away from the object. 

\subsection{The ADM mass}\label{sec:adm}
Let us point out that the full nonlinear solution to the $00$ Einstein equation can be expressed in terms of the ADM mass of the object, defined as
\be\label{defM}
M=-\int d^3x\,a^3\tau^0\,_0\,.
\ee
Let us demonstrate how it works in the subhorizon limit. Integrating both sides of the equation~\eqref{eq00lin} over some spatial region enclosing the object fixes the relationship between the fields $\Psi$ and $\pi$ in terms of the mass $M$. The solution to the equation \eqref{eqijlin} can be similarly expressed in terms of a volume integral over $\tau^i\,_i$. However, for stationary and virialized systems, as well as non-relativistic sources one can assume that the space integral of $\tau^i\,_i$ is either exactly zero, or is much smaller than the contributions from $\tau^0\,_0$ \cite{Hui:2010dn}. It can thus be neglected from the right-hand side of~\eqref{eqijlin}, giving the usual relationship $\Phi=\Psi$. 

Importantly, these observations hold for all types of objects:\\
$(i)$ For a \textit{test particle} all nonlinear contributions to $\tau ^\mu\,_\nu$ are negligible, so that $\tau^\mu\,_\nu=(\delta T^{\rm{m}})^\mu\,_\nu$. In particular, for a non-relativistic point mass located at the origin the only non-zero component of the stress tensor is $(\delta T^{\rm{m}})^0\,_0=-M_0\delta^{(3)}(0)/a(t)^3$  and the ADM mass \eqref{defM} coincides with $M_0$. In the case when the source is spherically symmetric and $\dot m_3=0$, one can solve for the field profiles, to get $\Phi,\Psi,\pi\sim -GM_0/(a(t)r)$. Strictly speaking, the rest mass $M_0$ coincides with the ADM mass only up to post-Newtonian corrections, \emph{i.e.} only up to terms coming from the nonlinear Einstein tensor in \eqref{deftau}. The size of these corrections can be estimated at the quadratic level to be of the order
\be
\frac{-\int d^3x\,\tau^0\,_0}{M_0}\sim \frac{\M^2\int d^3x\,(\Psi')^2}{M_0}\sim \frac{\frac{M_0^2}{\M^2}\frac{1}{r}}{M_0}\sim\frac{r_g}{r}\,,
\ee
where $r_g=M/\M^2$ is the Schwarzschild radius of the object. For a point particle, the contribution to the ADM mass from the non-linear stress tensor associated to the scalar field is negligible. We will discuss these in more detail in the context of motion of extended objects in Section~\ref{sec:violations}.
\\
$(ii)$ For an \textit{extended object}---an object of finite size, but negligible self-gravitational energy---the relevant contributions to $\tau^\mu\,_\nu$ are the first two terms in \eqref{deftau}. The nonlinear part of the Einstein tensor can still be neglected in this case.\\
$(iii)$ For a \textit{black hole}, there is no matter stress-tensor, \textit{i.e.} $\delta T^m_{\mu\nu}=0$. The pseudo-tensor $\tau^\mu\,_\nu$ is fully determined by the nonlinear contributions: both due to the gravitational self-interactions and those arising in the stress tensor of the scalar field. 

In any of the cases discussed, the equation \eqref{eq00lin} can be solved in terms of the ADM mass defined in \eqref{defM} while the right-hand side of Eq.~\eqref{eqijlin} can be neglected. The only unknown that remains is the solution to the scalar field equation \eqref{eqpilin}. This will be the subject of the next section.

\section{The (approximate) shift invariance of the scalar}\label{sec:shift}
The idea behind this section lies in the observation that, apart from the generic explicit $\phi$-dependence of the coefficients $\Lambda, c, H, m_2$ and $m_3$, the covariant action \eqref{act_cov} is invariant under constant shifts: $\phi(t,\vec x)\to\phi(t,\vec x)+\text{const}$. Moreover, since the unitary gauge action \eqref{act_unit} was introduced as an EFT of broken time translations during the dark energy dominated epoch of our universe, we are interested in a slowly varying Hubble rate and its derivatives. With this in mind, we can assume that also the other generically time dependent parameters in the action, $\Lambda(t),c(t),m_i(t)$, vary only mildly during one Hubble time. More specifically, we shall assume that $\dot m_i\sim Hm_i $ so that terms like $m_i\partial\pi\gg\dot m_i\pi$ in the subhorizon limit.\footnote{We note that the same reasoning would apply for the time dependence of the parameters in the EFT of inflation \cite{Cheung:2007st}.}  This would allow us to neglect all the terms in the equation of motion of the scalar field arising from the violation of the shift invariance and instead write it as a current conservation equation
\be
\nabla_\mu J^\mu = 0\,,
\ee
where $J^\mu$ is the current due to constant shifts of~$\phi$. 

\subsection{Static case}
The presence of the shift symmetry has proven to be very useful when discussing the equivalence principle in asymptotically flat spacetimes. It was used both for showing that the weak equivalence principle holds for theories with Vainshtein screening mechanism \cite{Hui:2009kc} and for the derivation of the no-hair theorem for static black holes in Galileon theories \cite{Hui:2012qt}. For time independent solutions, the divergence of the shift symmetry current in these cases can be reduced to $\nabla_\mu J^\mu=\partial_iJ^i$. 

The only shift symmetry breaking term contributing to the scalar fields equation of motion in these models is the scalars coupling to the trace of the matter stress-energy tensor, $\phi T^{\text{m}}\,_\mu\,^\mu$. Integrating this over some spatial volume and using the Gauss's law to convert it into a surface integral yields:
\be\label{eom_tens}
\oint dS_i\,J^i=-q\int d^3 x\,T^{\text{m}}\,_\mu\,^\mu\equiv qQ\,,
\ee
where $q = 1$ for ordinary matter and $q = 0$ for black holes. One can then choose the integration surface to lie far away from the moving object to get an asymptotic linear relationship between the fields $\Psi,\Phi,\pi$. Moreover, for objects with weak self-gravity the \emph{scalar charge} $Q$, defined in \eqref{eom_tens}, is given by the total inertial mass of the object, $Q\simeq M$ \cite{Hui:2010dn}. 
The scalar field profile, in the region asymptotically far from the object, can then be found to be of the form $\pi\sim -qGQ/r$, giving the interpretation of a scalar charge to the parameter $Q$. The non-vanishing field profile for $\pi$ is also referred to as the \textit{scalar hair}. Due to the couplings to the other gravitational potentials this introduces an additional gravitational force. The equivalence principle is then said to be obeyed in all cases when $Q$ only depends on the total inertial mass~$M$ and the constant $q$ does not vary from object to object. In this case, the modified gravity theory only redefines the effective Newton's constant. Importantly, in Galileon theories $q =0$ for black holes, due to the absence of the coupling $\phi T^{\text{m}}\,_\mu\,^\mu$. This phenomenon is known as the \emph{no-hair} theorem \cite{Hui:2012qt}. Hence, it is only the weak equivalence principle that is obeyed in flat space Galileon theories. 

Another terminology often used to make distinction of cases when the equivalence principle is violated is the concepts of \textit{primary} and \textit{secondary} hair~\cite{Coleman:1991ku}. The term `primary hair' is used when talking about a black hole hair that induces new quantum numbers. In the case of `secondary hair' the additional charge is entirely determined by the existing conserved charges, \textit{e.g.} like in the case when the scalar hair only depends on the ADM mass. 

Remarkable counterexamples are theories with Chameleon-type screening mechanisms, like $f(R)$ and Brans-Dicke theories. It was shown in~\cite{Hui:2009kc} that there the constant $q$ varies from $q=0$ for screened objects to $q = 1$ for unscreened objects, even when the gravitational self-interactions can be neglected. This manifests when considering the equation of the motion of the object: the difference in the scalar charge leads to a difference in the inertial and gravitational masses of the object. Hence, objects in such theories do not move on geodesics. 

\subsection{The current}\label{sec:currentfrw}
Let us analyze the extent up to which the shift symmetry is violated in the action \eqref{act_cov}. In general, given a covariant scalar field action of the form $S=\int d^4x\sqrt{-g}\,\mathcal L(\phi,\nabla_\mu\phi,\nabla_\mu\nabla_\nu\phi)$, the equation of motion for the scalar field can be put in the form
\be\label{eom_current}
{\mathcal E_\phi}\equiv\frac{1}{\sqrt{-g}}\frac{\delta S}{\delta \phi}=\frac{\partial\mathcal L}{\partial\phi}-\nabla_\mu J^\mu=0\, .
\ee
The current $J^\mu$ is defined as:
\be\label{current}
J^\mu=\frac{\partial\mathcal L}{\partial(\nabla_\mu\phi)}-\nabla_\nu\left(\frac{\partial\mathcal L}{\partial(\nabla_\mu\nabla_\nu\phi)}\right)\,
\ee
and is due to constant shifts of $\phi$; whereas a non-zero $\partial\mathcal L/\partial\phi$ only arises from the breaking of the shift symmetry in the action for $\phi$.\footnote{Let us for a moment discuss the difference between the shift symmetry in $\phi$ versus shift symmetry in $\pi$. Specifically, consider a situation when we had started from some covariant Lagrangian for $\phi$ admitting a more general homogeneous background solution $\phi_0(t)$ with $\dot \phi_0(t)\neq\text{ const}$. The unitary gauge action for perturbations around this solution would then correspond to setting $\phi(t,\vec x)=\phi_0(t)$. As in the previous case the invariance under time diffeomorphisms can be recovered by performing a transformation 
\be
t\to t+\pi(t,\vec x)\,,\qquad \phi_0(t)\to \phi_0(t)+\dot\phi_0(t)\pi(t,\vec x)\,.
\ee
However in the case when $\dot\phi_0\neq \text{const}$, the symmetry of $\phi$ under constant shifts does not imply a symmetry of $\pi$ under constant shifts. Indeed, under the shift $\phi\to\phi+c$ we get $\pi\to\pi+\dot\phi_0(t)^{-1}c$. Even though it is always possible to perform a field redefinition $\phi\to\tilde\phi(\phi)$ so that $\dot{\tilde\phi}_0=\text{const}$, this does not guarantee that the action for $\tilde\phi$ would remain shift invariant. Since it is the field $\pi$ that encodes the approximate invariance under the time translations then it is the shift invariance of $\pi$ that bears a physical importance. In the rest of this paper we shall assume that $\phi_0(t)=t$ and treat the shift symmetry of $\pi$ and $\phi$ as equivalent. 
}
 We give the explicit expressions of the current derived from the action \eqref{act_cov} in Appendix~\ref{app:current}.
 
 In the case when the shift invariance is only slightly broken, one would expect the equation of motion for the scalar to reduce to $\nabla_\mu J^\mu=0$ in the quasi-static limit. However, in our EFT of dark energy setup, it is precisely the explicit $\phi$~dependence of the action that drives the cosmological evolution. Hence $\partial\mathcal L/\partial\phi$ does contribute to the background evolution with
 \be\label{arg1}
 M_{\text{Pl}}^2\,\frac{\partial\mathcal L}{\partial\phi}=\dot c(t)-\dot\Lambda(t)\,.
\ee
 Another important difference from the current conservation in shift symmetric theories in the static case is that $J_0\neq 0$ on cosmological backgrounds. Indeed, the current components read:
  \be\label{arg2}
J_0=\frac{2c(t)}{\M^2}\,,\qquad J_i=0\,.
\ee
 Physically it makes sense---a cosmological background naturally picks out a time direction which consequently leads to a non-vanishing $J^0$. 
Combining \eqref{arg1} and \eqref{arg2} in the equation of motion~\eqref{eom_current} gives the standard conservation equation $\dot\rho_m+3H\rho_m=0$ on an FRW background. 

We face a more serious obstruction to the shift symmetry argument on FRW background when inspecting the higher order current components. 
As we show in Appendix~\ref{sec:current2}, the contribution to $\nabla_\mu J^\mu$ coming from $\nabla_0J^0$ is in general comparable to $\nabla_i J^i$. This, combined with the non-conservation of the current, $\partial\mathcal L/\partial\phi\neq 0$, prevents us from having an equation of the type \eqref{eom_tens} in a general case. In the absence of it, there is no advantage in recasting the scalar equation in the form~\eqref{eom_current}. Instead one has to work with the equation~\eqref{eqpilin} that can only be solved perturbatively and after integration contains an arbitrary integration constant that cannot be fixed by any guiding principle. One is thus unable to make exact non-perturbative arguments needed to discuss black hole hair and the strong equivalence principle. We therefore focus on extended objects in the remainder of this section.

\subsection{Extended objects in the subhorizon limit}\label{extended}

In theories with Vainshtein screening mechanism~\cite{Vainshtein:1972sx}, which is the case of all models described by the action \eqref{act_unit} (or, equivalently, \eqref{act_cov}), extended objects require a special attention. The reason for this is the presence of higher derivative scalar self-interactions, that, for an object of finite size, become dominant below some distance $r_V$ in the vicinity of the object. In these models the interactions responsible for the Vainshtein mechanism at the $n$th order in field perturbations are contained in terms of the type $(\partial\pi)^{2k}(\partial^2\pi)^{l}$ with $n=l+2k$, $k\geq1$. Due to the antisymmetric structure of these higher order self-interactions, in four spacetime dimensions only interactions up to $l=3$ appear in the most general Horndeski theories (see, for instance, \cite{Nicolis:2008in,Deffayet:2011gz}). Hence, in the scalar field equation of motion we expect that for $n=2,3,4$ the equation is dominated by $(\partial^2\pi)^n$ terms. 

For the choice of operators in the EFT action~\eqref{act_unit}, only the cubic Vainshtein interactions with $l=1$ are present giving a $(\partial^2\pi)^2$ term in the equation of motion at quadratic order. For extended objects with negligible gravitational self-interactions this is the leading contribution to the quadratic scalar fields equation of motion.

Due to the special derivative structure of the Vainshtein interactions, the equations of motion for perturbations around cosmological solutions on subhorizon scales always take the form of spatial divergences \cite{Kimura:2011dc}. At quadratic order, all the highest derivative contributions to both metric and scalar fields equations of motion take the form
 \begin{equation}\label{der_structure2}
 \begin{split}
 \Delta \mathcal X\Delta \mathcal Y-&\partial_i\partial_j\mathcal X\partial_i\partial_j\mathcal Y\\
 &=\partial_i\left(\partial_i \mathcal X\Delta \mathcal Y-\partial_j\mathcal X\partial_i\partial_j \mathcal Y\right)\,,
 \end{split}
 \end{equation}
 where $\{\mathcal X,\mathcal Y\}$ denote any of the fields $\{\Phi,\Psi,\pi\}$. 
For the expressions at higher orders, see \cite{Kimura:2011dc}. Indeed, we find that the quasi-static subhorizon contribution to the quadratic equation of motion of the scalar field reads:
 \be\label{divergence}
\mathcal E_{2,\pi}=-\M^2\partial_iJ^i=-\frac{m_3^3}{a^4}\partial_i\left[\partial_i\pi\Delta\pi-\partial_j\pi\partial_i\partial_j\pi\right]\,.
\ee
We also note that it is only the divergence of the spatial current component that contributes to the above equation. As we argue in Appendix~\ref{sec:current2}, both $\nabla_0J^0$ and $\partial\mathcal L/\partial\phi$ are subleading in comparison to the Vainshtein interactions at this order. 

At cubic order, only terms like $(\partial\pi)^2\partial^2\pi$ arise which, in the small field approximation, are negligible in comparison to the quadratic term. The same is true for all the other higher order terms that all can be safely neglected due to the fact that, by definition, gravitational self-interactions are negligible for extended objects and all the terms like $\pi,\partial\pi$ can still be treated perturbatively. Hence, the nonlinear equation of motion for $\pi$, for an extend object in subhorizon limit can be written as
\be\label{eomNL}
\mathcal{E}_{\text{NL},\pi}=\mathcal E_{2,\pi}=-\M^2\partial_i J^i\,.
\ee
Note that for later convenience we have included a factor of $\M^2$ in the definition of $\mathcal E_\pi$.

We can now run the argument, similar to that of the static case, for the current conservation for extended objects. In particular, by substituting~\eqref{eomNL} on the right-hand side of the equation \eqref{eqpilin} we see that one can reduce the volume integral to a surface integral. This is true for arbitrary location of the surface, even inside the Vainshtein radius. Choosing the surface to lie far away from the object, \emph{i.e.} for $r\gg r_V$, allows one to treat $\pi$ linearly. In spherically symmetric case this gives the asymptotic relation:
\be
\partial_r\bigg[m_3^3\Phi+\left(2c+(m_3^3)^{\hbox{$\cdot$}}+Hm_3^3 \right)\pi\bigg]=0
\ee
In combination with \eqref{eq00lin} and \eqref{eqijlin} this allows one to solve for all the gravitational field profiles solely in terms of the ADM mass. The above solution is only valid for the gravitational field far away from the object, at distances that exceed its Vainshtein radius. Nevertheless there is no fundamental obstacle in determining the gravitational fields also at shorter distances by using the non-linear expression \eqref{eomNL}. These would then have to match the asymptotic solutions found above which, as we have shown, do not contain any new additional charge and are fixed by the ADM mass of the object. Hence the weak equivalence principle holds up to post-Newtonian and \emph{post-Vainshtein} corrections. We shall discuss what we mean by the latter in Section~\ref{sec:violations} where we shall study the motion of the extended object in an \emph{external} gravitational field.

\section{Equivalence principle in de Sitter universe}\label{sec:equiv_dS}
An outstanding example of where the vanishing of the current is an exact statement is the case of de Sitter spacetime. For instance, there are known numerical black hole solutions with de Sitter asymptotics in the cubic Galileon theory \cite{Babichev:2016fbg}. These solutions involve a time dependent scalar field profile and a static solution for the metric, written in the static coordinate patch of de Sitter. In general, their $J^0\neq 0$ while $J^\rho = 0$ coincides with the $0\rho$ component of the Einstein equation (where $\rho= re^{Ht}$ is the radial coordinate in the static de Sitter coordinate system\footnote{The notations for the static coordinates $\rho, \tau$ and the FRW coordinates $r,t$ are exchanged in \cite{Babichev:2016fbg}.}). The fact that a time dependent scalar field profile can be compatible with the vanishing of the spherical component of the Noether current, allowing for static and spherically symmetric metric solutions was already noticed in \cite{Babichev:2015rva}. The findings of \cite{Babichev:2016fbg} also show that there is a non-vanishing scalar field profile associated with their black hole solutions. At small scales the scalar field solution depends on an arbitrary integration constant that can be interpreted as the primary hair of the black hole. The large scale asymptotics however only depend on the black hole mass and thus only give a secondary hair. Overall, it seems to indicate that $(i)$~the black holes do have hair on cosmological backgrounds, even in shift symmetric theories, and that $(ii)$~the strong equivalence principle \textit{might} hold for black holes on de Sitter space. 

We shall now show that the equivalence principle does indeed hold for spherically symmetric solutions of the action \eqref{act_cov} in de Sitter space. Restricting to de Sitter universe in our EFT implies setting $\rho_{\text m}=p_{\text{m}}=c = 0$ and $\Lambda = 3\M^2H^2 = \text{const}$.  Inspired by the ansatz of \cite{Babichev:2016fbg} we shall assume that the scalar fields $\Phi,\Psi,\pi$ are not only spherically symmetric, but also that the dependence on $t,r$ only comes through 
\begin{equation}\label{symmetry}
\Phi(r,t) = \Phi(re^{Ht})=\Phi(\rho)\,\quad\text{etc.}
\end{equation}
That it is reasonable to ask for such static solutions to be compatible with the expansion of the spacetime is due to the isometries of de Sitter space. Indeed, the metric \eqref{gauge} on the background with $a(t)=e^{Ht}$, rewritten in spherically symmetric coordinates is invariant under $t\to t+\lambda$, $r\to e^{-H\lambda}$. Since under this transformation $\rho=re^{Ht}\to \rho$ is also invariant, then the static ansatz \eqref{symmetry} is compatible with the isometries of de Sitter background.\footnote{We thank Andrew Tolley for pointing this out.}
In fact, it also has a physically well motivated meaning since $\rho=re^{Ht} = a(t) r$ is just the physical radius as opposed to the comoving one. For instance, also the gravitational potential in GR around a spherically symmetric static source in an FRW universe equals to $\Phi=-GM/(a(t)r)$. We also emphasize that the static ansatz \eqref{symmetry} can only be a valid solution if also the coefficients $m_2$ and $m_3$ are time independent. Therefore, in the remainder of this section we shall set 
\be
m_2\,,m_3=\text{const}\,.
\ee
It might appear that since the time diffeomorphisms (including constant shifts) are broken by our EFT construction, forbidding these coefficients to bear any time dependence is unnatural. Nevertheless it is justified when constraining the analysis to de Sitter spacetime. Time dependence of $m_2,m_3$ in this case introduce deviations from exact de Sitter and their time derivatives can be regarded as additional slow-roll parameters. We shall come back to this point in Section~\ref{sec:generalize}.

The linearized equations of motion of the field perturbations for the vacuum configuration (\emph{i.e.} in the absence of any additional matter sources besides the metric and the scalar field) can be simplified to:
\begin{align}
\Psi' - \Phi'=0\,,\quad\Phi+\rho\,\Phi'=0\,,\nonumber\\\label{lineomdS}
-\Phi+H\rho\,\pi'=0\,,
\end{align}
where the prime denotes a derivative with respect to the physical radius $\rho$. The above equations are solved by $1/\rho$ profiles for all the three fields. A few remarks are in order. In particular, let us note that the first equation in \eqref{lineomdS} follows from the off-diagonal $ij$ components of the Einstein equations and has been integrated over the radial coordinate. On the other hand, the second and third equation in \eqref{lineomdS} are specific combinations of the $tx$ and $xx$ components of the Einstein equations and only depend on the first order derivatives without the need of integration. This already points towards the fact that the asympotic relationships between the three fields $\Phi, \Psi, \pi$ are independent on the nonlinearities of the system under the symmetry conditions \eqref{symmetry}. 

\subsection{The current}
Let us elaborate on this statement by looking at the current components as defined in~\eqref{current}. In particular, we will show that the asymptotic relationship \eqref{lineomdS} can be expressed as the vanishing of one of the current components. Due to the field dependence \eqref{symmetry} it proves useful to switch from the Cartesian coordinate system first to the spherical coordinates and then to the coordinate system $(\tau,\rho,\varphi,\theta)$ with $\tau = t$ and $\rho = a(t)r$. The metric~\eqref{gauge} thus becomes:
\begin{align}\label{metric2}
ds^2 &= \left(-e^{2\Phi(\rho)}+H^2\rho^2e^{-2\Psi(\rho)}\right)d\tau^2+\\
&-2H\rho \,d\rho d\tau +e^{-2\Psi(\rho)}d\rho^2+\rho^2e^{-2\Psi(\rho)}\,d\Omega^2\,.\nonumber
\end{align}
We then find the $J^\rho$ component of the current from Eq.~\eqref{current} by first finding the current components $J^t, J^i$ in the comoving Cartesian coordinates and then transforming them as
\begin{align}\label{jrho}
J^\rho & = a(t)^2\frac{x^iJ^i}{\rho}+\rho HJ^t\\
&=4H\rho \,m_2^4\left(-\Phi+H\rho\pi'\right)\nonumber\\
&-3H^2\rho \,m_3^3\left(\Phi+\rho\Psi'\right)
-{m_3^3}\left(H\,\pi'+\Phi'\right)\,.\nonumber
\end{align}
By comparing the above expression to the equations of motion \eqref{lineomdS} it is straightforward to see that $J^\rho = 0$ on shell. Hence, the radial component of the current vanishes exactly on the equations of motion. Curiously, also the linear $J^\tau$ component vanishes on shell. Indeed, 
\begin{align}
J^\tau&=4m_2^4\left(-\Phi+H\rho\pi'\right)\nonumber\\
&-3H \,m_3^3\left(\Phi+\rho\Psi'\right)
-\frac{m_3^3}{\rho^2}\left(2\rho\,\pi'+\rho^2\pi''\right)\,.\nonumber
\end{align}
The vanishing of the radial current component can be understood when considering the variation of the action \eqref{act_cov} under diffeomorphism transformations. This relation was observed earlier in \cite{Babichev:2015rva} for static black holes; we shall follow their analysis closely below. 

A variation of the action \eqref{act_cov} can be written in a general form as
\begin{equation}
\delta S = \int d^4 x\,\sqrt{-g}\left(\frac{\delta S}{\delta g^{\mu\nu}}\delta g^{\mu\nu}+\frac{\delta S}{\delta \phi}\delta\phi\right)\,.
\end{equation}
If the variation is due to a diffeomorphism transformation
\begin{equation}
x^\mu\quad\to \quad\tilde x^\mu=x^\mu+\xi^\mu(x)
\end{equation}
the change in the metric and the scalar field is accordingly:
\begin{equation}
\delta g^{\mu\nu}=\xi^{\mu;\nu}+\xi^{\nu;\mu}\,,\quad \delta \phi=-\partial_\mu\phi\,\xi^\mu\,,
\end{equation}
where $\xi^{\mu;\nu}\equiv g^{\nu\alpha}\nabla_\alpha\xi^\mu$. By defining the metric equations of motion as
\begin{equation}
\mathcal E_{\mu\nu}\equiv\frac{2}{\sqrt{-g}}\frac{\delta S}{\delta g^{\mu\nu}}\,
\end{equation}
and the scalar field equation of motion $\mathcal E_\phi$ as in~\eqref{eom_current} we can rewrite the variation as
\begin{align}\label{var_gen}
\delta S &= \int d^4 x\,\sqrt{-g}\left[\frac{1}{2}\mathcal E_{\mu\nu}\left(\xi^{\mu;\nu}+\xi^{\nu;\mu}\right)\right.\nonumber\\
&\left.+\left(\nabla_\alpha J^\alpha-\frac{\partial\mathcal L}{\partial\phi}\right)\partial_\mu\phi\,\xi^\mu\right]\,.
\end{align}
The traditional use of this expression is to obtain the conservation equations (the so-called \emph{Bianchi identities}):
\begin{equation}\label{bianchi}
\nabla^\nu\mathcal E_{\mu\nu}+\mathcal E_\phi\partial_\mu\phi=0\,.
\end{equation}
This equation has a twofold significance. First, it means that on scalar fields equation of motion the stress-energy tensor is conserved. Second, it shows that the equations of motion of the metric and of the scalar field are not all independent, but are related through \eqref{bianchi}. Indeed, for our field configuration \eqref{phi}, \eqref{symmetry}, and \eqref{metric2}, its 0-th component at linear level reads
\begin{equation}\label{offshell}
\partial_\tau\left(\mathcal E^\tau_\tau-M_{\text{Pl}}^2\,J^\tau\right)+\frac{1}{\rho^2}\partial_\rho\left(\rho^2\left(\mathcal E^\rho_\tau- M_{\text{Pl}}^2\,J^\rho\right)\right)=0\,.
\end{equation}
This is an off-shell equation and is exactly satisfied.

There is yet another way how to make use of the variation of the action above. To see this let us consider a diffeomorphism 
\begin{equation}\label{ximu}
\xi^\mu(x)=\left(\xi^\tau(\rho),0,0,0\right)\,.
\end{equation}
One can show that for this choice of $\xi^\mu$ we have
\begin{equation}\label{xidS}
\mathcal E_{\mu\nu}\left(\xi^{\mu;\nu}+\xi^{\nu;\mu}\right)=2\mathcal E_\tau^\rho\,\partial_\rho\xi^\tau\,,
\end{equation}
valid up to arbitrary order in perturbations. For the variation of the action we thus obtain
\begin{align}\label{var}
\delta S = \int d^4x\,\sqrt{-g}\left[\partial_\rho\xi^\tau\left(\mathcal E_\tau^\rho-M_{\text{Pl}}^2\,J^\rho\right)-\frac{\partial\mathcal L}{\partial\phi}M_{\text{Pl}}^2\,\xi^\tau\right]\,.\end{align}
Hence, in the absence of an explicit violation of the shift invariance, \textit{i.e.} when $\partial\mathcal L/\partial\phi=0$, one arrives at the relationship:\footnote{We should note that choosing the gauge parameter $\xi^\mu$ to be time independent makes the variational problem somewhat subtle. In general, the relationship \eqref{epsrhotau} should come with an additional time integral $\int_{\tau_{\text{initial}}}^{\tau_{\text{final}}}d\tau$\,. However, in an entirely static in $\tau$ situation, which in our case is equivalent to saying that $\partial\mathcal L/\partial\phi=0$, the integral is trivial. We thank Alberto Nicolis for extended discussions on this.}
\begin{align}\label{epsrhotau}
&\mathcal E_\tau^\rho-M_{\text{Pl}}^2\,J^\rho=0\,.
\end{align}
Given the expression \eqref{non_cons}, we see that this is precisely the case in de Sitter universe with time independent coefficients $m_2,m_3$. From the above equation it thus follows that the radial component of the current coincides with the equation of motion $\mathcal E_\tau^\rho$ at all orders. One can check by explicit computations at linear order that indeed the equation~\eqref{epsrhotau} is satisfied off-shell. 

Hence, we conclude that the radial component of the current vanishes on shell at all orders in perturbations:
\begin{equation}\label{jzero}
J^\rho = 0\,.
\end{equation}
In particular, the linearized version of the above equation gives an asymptotic relationship between the fields $\Phi,\Psi,\pi$ and their first derivatives. 

\subsection{Nonlinearities, matter and black holes}\label{sec:solution}
The linearized equations \eqref{lineomdS} are valid in vacuum and do not tell what the gravitational fields around an object of a given mass are. To take this into account, let us retain the pseudo stress tensor $\tau^0_0$, as defined in \eqref{deftau}, in our equations of motion. We note that the variation \eqref{var} include $\tau^0_0$ implicitly, since $\mathcal E_{\mu\nu}$ refers to the full metric equations of motion. Hence, also the statement about the vanishing of the current \eqref{jzero} remains unchanged and gives one equation for the three fields $\Phi,\Psi,\pi$.

If $\tau^0_0$ is only due to nonlinearities, \textit{i.e.} $\delta T^{\text{m}}_{\mu\nu}=0$, then we are considering black holes. In turn, a point-like test particle placed at the origin would have $\tau^0_0=-M\delta^{(3)}(0)/a(\tau)^3$. In the following we shall leave the internal decomposition of $\tau^0_0$ unspecified and solve for the gravitational fields sourced by it. 

Due to the spherical symmetry of the system, the only non-zero components of the metric equations of motion are $\mathcal E_{\tau\tau}\,,\mathcal E_{\tau\rho}\,,\mathcal E_{\rho\rho}\,,\mathcal E_{\varphi\varphi}\,,\mathcal E_{\theta\theta}$. Moreover, the last three of them only depend on two independent combinations of the orginial Cartesian equations of motion, defined through:
\begin{align}
\mathcal E_{ij}&\equiv\mathcal E_{ij}^{\text{off}}+\delta_{ij}\mathcal E^{\text{diag}}\,,\\
 \mathcal E_{ij}^{\text{off}}&\equiv\left(x_ix_j-\frac{1}{3}\delta_{ij}r^2\right)\mathcal E^{\text{off}}\,.
\end{align}
Combined with the equation of motion for the scalar and with the $\tau$ and $\rho$ components of the Bianchi identities \eqref{bianchi} we have in total $4+1-2=3$ equations for the three fields $\Psi,\Phi,\pi$. The most convenient choice is setting $J^\rho,\,\mathcal E_{\tau\tau},\,\mathcal E^{\text{off}}$ to zero. As in the case of the subhorizon limit, the off-diagonal equation allows us to set $\Phi=\Psi$. Using this, vanishing of $J^\rho$ gives the following relationship:
\begin{equation}\label{phipi}
-\Phi+H\rho\pi'=m_3^3\frac{1+3H^2\rho^2}{-m_3^3+4H\rho^2m_2^4}(\rho\Phi)'\,.
\end{equation}
We note that the right-hand side of this equation vanishes for $1/\rho$ profiles, \textit{i.e.} when $\rho\Phi=\text{const}$. Finally, the $\mathcal E_{\tau\tau}$ equation reads:
\begin{align}
0&=m_3^3\Delta\pi+2M_{\text{Pl}}^2\Delta\Phi+\tau^0_0\\
&+\left(4m_2^4+3m_3^3H\right)\left(\Phi-H\rho\pi'\right)\nonumber\\
&+3H\left(m_3^3-2HM_{\text{Pl}}^2\right)(\rho\Phi)'\,.\nonumber
\end{align}
By using the equation \eqref{phipi}, the second line can be expressed in terms of $(\rho\Phi)'$ and thus only the first line survives for $1/\rho$ profiles. It can be solved in terms of the ADM mass, defined in \eqref{defM}, to give:
\begin{equation}\label{sol_ep}
\Phi+\frac{m_3^3}{2M_{\text{Pl}}^2}\pi=-\frac{MG}{\rho}\,,
\end{equation}
where $G$ is the Newton's constant with $2M_{\text{Pl}}^2=(4\pi G)^{-1}$\,. Together with Eq.~\eqref{phipi} and the relation $\Phi=\Psi$ this determines both the scalar field profile and the gravitational potential felt by the object entirely in terms of its ADM mass. This means that the equivalence principle is obeyed in these theories, since the ADM mass bears no distinction between how it is split into rest masses, energy from self-interactions or self-gravitational energy. Hence, it treats all objects on equal footing.

We also observe that it is obvious from the equation above that a time dependent $m_3(\tau)$ would be inconsistent with the solution \eqref{sol_ep} for a  conserved ADM mass~\eqref{defM}. Therefore, for the field ansatz~\eqref{symmetry} to be a solution of our system, we must restrict our argument to time independent $m_3$, as argued earlier. At the same time, a time dependent $m_2$ still allows for static gravitational potentials at linear level. Indeed, for $1/\rho$ profiles, the time dependence of $m_2$ would be irrelevant for solving Eq.~\eqref{phipi}. However, at higher orders $\dot m_2\neq 0$ would inevitably introduce time dependence in the system and thus must be set to zero for field configurations \eqref{symmetry}. We therefore conclude that time-dependent $m_2,m_3$ violate the equivalence principle. We discuss it together with the violation due to the time dependence of the Hubble parameter in the next section. 

Finally, let us point out that the solution given by equations \eqref{phipi} and \eqref{sol_ep} is a valid spherically symmetric, static (in physical coordinates) star and black hole solution in de Sitter space. Transforming the coordinates $(\tau,\rho)$ to the static patch coordinates $(\tilde\tau,\rho)$ by
\be
\tilde \tau=\tau-\frac{1}{2H}\ln \left(1-H^2\rho^2\right)\,,
\ee
we obtain
\begin{align}
&\tilde g_{\tau\tau}=-\e^{2\Phi}+H^2\rho^2\e^{-2\Psi}\,,\\
&\tilde g_{\rho\rho}=\frac{\e^{-2\Psi}-H^2\rho^2\e^{2\Phi}}{(1-H^2\rho^2)^2}\,,\\
&\tilde g_{\tau\rho}=\frac{ H\rho}{1-H^2\rho^2}\left(\e^{2\Phi}-\e^{-2\Psi}\right)\,.
\end{align}
In the subhorizon small field limit with $H\rho,\Phi\ll1$ it reduces to the Schwarzschild-de Sitter metric with $f(\rho)=1+2\Phi-H^2\rho^2$.

\subsection{Generalize to FRW?}\label{sec:generalize}
An FRW universe with a slowly changing Hubble rate can be described by the theory \eqref{act_cov} upon including a background stress-energy tensor $T^{\rm m}\,^\mu_\nu=\text{diag}(-\rho_{\text{m}},p_{\text{m}},p_{\text{m}},p_{\text{m}})$. In the following we shall assume a pressureless matter with $p_{\text{m}}=0$. As opposed to the case of de Sitter, now the matter energy density is time dependent and thus, also the Hubble rate and the parameters $\Lambda=\Lambda(\tau)$ and $c=c(\tau)$, as determined by equations~\eqref{backeqcL}. 

The complexity of the equations of motion increases considerably and we cannot obtain an analytic solution even in the absence of any additional matter source and nonlinearities. While the solution can in principle be found numerically, the previous argument that allowed us to set $J^\rho=0$ in the case of de Sitter cannot be applied anymore. Let us demonstrate how several of the steps in derivation that lead to \eqref{jzero} cannot be performed anymore. 

As before, we wish to consider variation of the action \eqref{var_gen} under a diffeomorphism \eqref{ximu}. For a time dependent system we have to allow for $\xi^\tau=\xi^\tau(\tau,\rho)$. We then find
\begin{align}
\frac{1}{2}\mathcal E_{\mu\nu}&\left(\xi^{\mu;\nu}+\xi^{\nu;\mu}\right)=\nonumber\\
&\mathcal E^\tau_\tau\,\partial_\tau\xi^\tau+\mathcal E_\tau^\rho\,\partial_\rho\xi^\tau-\rho\dot H\mathcal E^\tau_\rho\xi^\tau\,.
\end{align}
In comparison to \eqref{xidS} there is an extra term proportional to $\xi^\tau$. Upon inserting this in the variation of the action we obtain
\begin{align}\label{varfrw}
\delta S = &\int d^4x\,\sqrt{-g}\left[\partial_\rho\xi^\tau\left(\mathcal E_\tau^\rho-M_{\text{Pl}}^2\,J^\rho\right)\right.\\
&\left.+\partial_\tau\xi^\tau\left(\mathcal E_\tau^\tau-M_{\text{Pl}}^2\,J^\tau\right)\right.\nonumber\\
&\left.-\left(\rho\dot H\mathcal E^\tau_\rho+\frac{\partial\mathcal L}{\partial\phi}M_{\text{Pl}}^2\right)\,\xi^\tau\right]\,.\nonumber
\end{align}
We further note that in the presence of $\rho_{\text{m}}$ the background values of $J^\mu$ and $\partial\mathcal L/\partial\phi$ become
\begin{align}
&J^\tau=\frac{2c(\tau)}{M_{\text{Pl}}^2}\,,\quad J^\rho=H\rho J^\tau\,,\\
&M_{\text{Pl}}^2\,\frac{\partial\mathcal L}{\partial\phi}=\dot c(\tau)-\dot\Lambda(\tau)\,,
\end{align}
giving additional contributions when expanding the variation \eqref{varfrw} up to given order in perturbations. Moreover, in distinction from de Sitter case, $\partial\mathcal L/\partial\phi$ is non-zero already at linear level, as is obvious from the non-perturbative expression \eqref{non_cons}. The resulting conservation equation of our interest can then be obtained by varying the action with respect to $\xi^\tau=\xi^\tau(\tau,\rho)$ and gives:
\begin{align}\label{eom_frw}
-\partial_\rho&\left(\sqrt{-g}(\mathcal E_\tau^\rho-M_{\text{Pl}}^2\,J^\rho)\right)\nonumber\\
&-\sqrt{-g}\left(\rho\dot H\mathcal E^\tau_\rho+M_{\text{Pl}}^2\frac{\partial\mathcal L}{\partial\phi}\right)\nonumber\\
&-\partial_\tau\left(\sqrt{-g}\left(\mathcal E^\tau_\tau-\M^2J^\tau\right)\right)=0\,.
\end{align}
This is again an off-shell equation. In distinction from \eqref{epsrhotau} the first line appears with a radial derivative and there are new contributions due to the time dependence. 
In particular, there is no way of arguing that the term in the second line can be dropped. Moreover, one can explicitly check that it contains contributions of the same order as the term in the first line. Indeed, at linear oder we find:
\begin{align*}
(1^{\text{st}})&\supset4\rho^3H^3m_3^3\left(\frac{\dot H}{H^2}+\frac{(m_3^3)^{\hbox{$\cdot$}}}{Hm_3^3}\right)\pi'\\
(2^{\text{nd}})&\supset\rho^3H^2\left[\rho_{\text{m}}\left(-3+\frac{\dot H}{H^2}\right)\right.\\
&\left.+4m_3^3\frac{\dot H}{H}+2M_{\text{Pl}}^2H^2\left(\left(\frac{\dot H}{H^2}\right)^2+\frac{\ddot H}{H^3}\right)\right]\pi'\,,
\end{align*}
where we use the notations $(1^{\text{st}})$ and $(2^{\text{nd}})$ for the first and second line in \eqref{eom_frw} respectively. We note that all the terms appear due to the presence of matter, $\rho_{\text{m}}$, and the time dependence of the Hubble parameter and $m_3$. Importantly, we see that the $m_3^3$ terms give contributions of the same order. 

Thus one cannot neglect the impact of the FRW spacetime and there is no way to argue that an equation of type \eqref{jzero} should exist. Nevertheless, one should keep in mind that in the context of dark energy (or, similiarly, of inflation) one is interested in quasi de Sitter spacetimes. Thus one can work with the result of de Sitter, where we have shown that the equivalence principle holds, and only consider violations of it due to the time dependence of the Hubble rate and parameters $m_2,m_3$. In particular, one could introduce a set of approximately constant parameters corresponding to 
\begin{equation}
\rho_{\rm m}\,, \frac{\dot H}{H^2}\,,\frac{\ddot H}{H^3}\,,\frac{(m_2^4)^{\hbox{$\cdot$}}}{Hm_2^4}\,,\frac{(m_3^3)^{\hbox{$\cdot$}}}{Hm_3^3}\,,\dots
\end{equation}
to quantify these deviations. It is beyond the scope of the present paper and will be left for future work.

\section{Geodesics in FRW}\label{sec:geodesics}
In this section we shall compute the acceleration felt by an object moving in an external configuration of the fields $\{\Phi_0,\Psi_0,\pi_0\}$ in the modified gravity theories \eqref{act_unit}. We shall then check whether the equivalence principle holds in these models, \emph{i.e.} whether the object moves on a geodesic. A particular emphasis will be put on describing the motion of finite size (extended) objects. For this we shall use the approach of Einstein, Infeld and Hoffmann \cite{Einstein:1938yz,damour} and will follow closely the procedure of \cite{Hui:2009kc}. 

An alternative formalism for describing the dynamics of non-relativistic extended objects coupled to gravity relies on using effective field theory methods \cite{Goldberger:2004jt,Goldberger:2007hy}. In flat space, one can see gravity as a theory of an interacting spin-two particle described by a tensor field $h_{\mu\nu}\equiv g_{\mu\nu}-\eta_{\mu\nu}$. The extended object can in turn be approximated as a point-particle with a worldline coordinate $x^{\mu}(\lambda)$ that depends on an arbitrary affine parameter $\lambda$. It is possible to describe their mutual interactions, as well as their interactions with external fields, by constructing an EFT Lagrangian that is consistent with the symmetries of the long-wavelength physics. In our case the relevant symmetry is the general coordinate invariance. At leading order, the effective action describing the graviton and a point particle of mass $M$ is simply:
\be\label{eff_pp}
\begin{split}
S_{\rm eff}&=S_{\rm gravity}+S_{\rm pp}\\
&=\frac{\M^2}{2}\int d^4x\sqrt{-g}\,R-M\int d\tau\,,
\end{split}
\ee
where $d\tau^2=g_{\mu\nu}(x(\lambda))dx^\mu dx^\nu$ is the proper time variable. Varying with respect to $x^\mu$ gives the usual geodesic equation of a test particle. The corrections due to the finite size effects then emerge from the higher order operators that can be added to the point particle action \cite{Goldberger:2004jt,Goldberger:2007hy}: 
\be
S_{\rm pp}=-M\int d\tau+c_R\int d\tau R+c_V\int d\tau\,R_{\mu\nu} \dot x^\mu\dot x^\nu+\dots\,,
\ee
where $\dot x^\mu=dx^\mu/d\tau$ and the dots stand for an infinite series of higher curvature corrections. This EFT is valid at energy scales $kr_s\ll 1$ where $r_s$ is the size of the object. The precise values of the coefficients $c_V, c_R$ can be obtained from a UV matching procedure to some microscopic model of the internal structure of the object. The idea is then to compare the observables computed in the full short distance theory and in the effective theory.

In the case when gravity is modified to a scalar-tensor theory, the additional scalar degree of freedom, $\pi$, also needs to be included in the EFT description \eqref{eff_pp}. In particular, in a shift-invariant theory with Vainshtein screening mechanism the effective action describing the coupling of the scalar field to a localized source would include operators like
\be
\begin{split}
\int d\tau\,\left(c^{(1)}\partial_\mu\partial_\nu\pi+c^{(2)}\partial_\mu\partial_\alpha\pi\partial^\alpha\partial_\nu\pi+\dots\right)\dot x^\mu\dot x^\nu\,.
\end{split}
\ee
For an extended object, the dimensionful coefficients $c^{(1)},c^{(2)}$ parametrize our ignorance about the internal structure and finite-size effects of the object. Importantly, inside the Vainshtein region the second order derivatives $\partial^2\pi$ can lead to sizeable effects. The point that we want to emphasize here is that \emph{a~priori} one would therefore expect  these operators to modify the geodesic equation of the extended object. The upshot of this section is to show, in an indirect way, that such corrections do not arise in the scalar-tensor theories governed by the action~\eqref{act_cov}. 

\subsection{The method}\label{sec:method}
In what follows we shall be working in the Cartesian FRW coordinates \eqref{gauge} and will always take the quasi-static subhorizon limit.
To characterize the object, in addition to its mass $M$, defined in \eqref{defM}, we also define its position and momentum:
\begin{align}
X^i&\equiv -\int d^3x\,a^3\,x^i \,\tau_0\,^0/M\,\label{defXi}\,,\\
 P_i&\equiv \int d^3 x\, a^3 \,\tau_i\,^0\label{defPi}\,,
\end{align}
where the volume integral is taken over a sphere of radius $r$ enclosing it. The gravitational force exerted on the moving object can then be found as the time derivative of the momentum $\dot P_i$. In order to test the equivalence principle we shall relate the force to the acceleration of the object, calculated as $\ddot X^i$. In order to take the time derivatives of $P_i$ and $X_i$ and to manipulate the volume integrals in the definitions \eqref{defXi} and \eqref{defPi}, we will need various conservation laws that we present below.

\subsubsection{Conservation laws}
The Bianchi identity of the Einstein tensor, i.e. the fact that $\nabla_\mu G^\mu\,_\nu=0$, guarantees that the linearized Einstein tensor obeys the following linearized conservation laws:
\begin{align}
&\partial_\nu\left(a^3G^{\rm{L}}\,_0\,^\nu\right)=a^3HG^{\rm{L}}\,_i\,^i\,,\\
&\partial_\nu\left(a^3G^{\rm{L}}\,_i\,^\nu\right)=a^3\partial_i\Phi\left(\bar G^0_0-\frac{1}{3}\bar G^i_i\right)\,,
\end{align}
where the bar denotes the background values. We remark that the first relation is only valid in the subhorizon limit. Combining the above expressions with the Einstein equations \eqref{Einsteineqs} and the covariant conservation law \eqref{eqpi} we can derive the conservation laws for the pseudo stress-energy tensor $\tau^\mu\,_\nu$. These read:
\begin{align}\label{cons0}
&\partial_\mu\left(a^3\tau^\mu\,_0\right)=a^3\left[H\tau^i\,_i+\mathcal E_{\rm{NL},\pi}\right]\\\label{consi}
&\partial_\mu\left(a^3\tau^\mu\,_i\right)=-a^3\partial_i\Phi(\rho_{\text{m}}+p_{\text{m}})\,,
\end{align}
where $\mathcal E_{\rm{NL},\pi}$ is defined as the nonlinear part of Eq.~\eqref{eqpilin}. We stress again that the first relation in only valid in the subhorizon limit while the second equation is exact. We see that the $0th$ component of the pseudo stress tensor is not conserved at nonlinear level. In comparison, for an object moving on a flat background in GR, the right-hand side of the first equation vanishes. 

Finally, let us note that due to \eqref{cons0} in general also the mass $M$ in \eqref{defM} is not conserved in theories~\eqref{act_unit}. Instead, we obtain
\be\label{dotM}
\dot M=\int d^3x\,a^3\left[-\mathcal E_{\rm{NL},\pi}-H\tau^i\,_i+\partial_i\tau^i\,_0\right]\,.
\ee
As was explained in \ref{sec:adm}, the integral over the second term vanishes for stationary and virialized systems. The last term can in turn be transformed into a surface integral $\oint dS_i\,\tau^i\,_0$ that vanishes under the assumption of zero energy flux through the integration surface. Hence, the non-conservation of the ADM mass, only arises due to the first term in \eqref{dotM}. 

\subsubsection{Geodesic equation}
In order to write the geodesic equation, we will need to relate the force $\dot P_i$ to the acceleration of the object $\ddot X^i$. Let us first compute $\dot X^i$. Using the defenition~\eqref{defXi} together with the conservation law \eqref{cons0} we can write
\begin{align}\label{dotXi}
\dot X^i=&\frac{1}{a^2}\frac{P_i}{M}\\
&-\frac{1}{M}\int d^3x\,x^i\,a^3\left[H\tau^j\,_j+\mathcal E_{\rm{NL},\pi}\right]-\frac{\dot M}{M}X^i\,.\nonumber
\end{align}
In deriving the above expression we have neglected the surface integral $\int dS_j\,x^i\tau^j\,_0$. For future reference let us define the integral appearing in the above equation as:
\be\label{nonlinear}
\mathcal I(t)\equiv\int d^3x\,x^i\,a^3\left[H\tau^j\,_j+\mathcal E_{\rm{NL},\pi}\right]\,.
\ee
By taking another derivative we arrive to the expression for the acceleration 
\begin{align}\label{geodesic}
\ddot X+2H\dot X=\frac{1}{a^2M}\partial_0\left( P_i
-a^2\,\mathcal I\right)\,,
\end{align}
where we have omitted the terms arising due to the non-conservation of the ADM mass for reasons that we shall explain below. We note that the additional Hubble friction term on the left-hand side of this equation arises naturally in the geodesic equation for a worldline $X^\mu(\lambda)$ in FRW spacetime, which in the absence of the force would read $\frac{d^2X^\mu}{d\lambda^2}+\Gamma^\mu_{\nu\rho}\frac{dX^\nu}{d\lambda}\frac{dX^\rho}{d\lambda}=0\,$.

The quantity on the right-hand side of \eqref{geodesic} gives the force exerted on the moving object. For this we need to evaluate the time derivative of the momentum, defined in~\eqref{defPi}. By using the conservation law~\eqref{consi} we find
\be\label{dotPi}
\dot P_i=-a^3\oint dS_j \tau^j\,_i -a^3\int d^3 x\,(\rho_{\text{m}}+p_{\text{m}})\,\partial_i\Phi\,,
\ee
where the integration measure $dS_j=dA \cdot n_j$ denotes a surface integration with $dA$ being the surface area element with the outgoing unit normal $n_j$. The integration surface can be chosen so that $\delta T^{\text{m}}\,_\mu\,^\nu$ vanishes on it so that one only needs to evaluate the nonlinear contributions to the integrand $\tau^j\,_i$. The advantage of defining the momentum $P_i$ through the pseudo stress tensor $\tau_\mu\,^\nu$ defined as in \eqref{deftau} is apparent now---according to \eqref{consi} it is conserved and thus we were able to reduce the computation of its time derivative to a surface integral instead of a volume integral.  This allows us to compute the gravitational force felt by the moving object without specifying its internal structure. Moreover, we only need to specify the fields on the boundary, which we can choose to be far away from the source, where one can use the perturbative expansion.

\subsubsection{Background--object split}
In order to evaluate the surface integral, we decompose the field perturbations $\mathcal X=\{\Phi,\Psi,\pi\}$ on the surface of the sphere, centered around $r=0$ and enclosing the object of interest, in two parts~\cite{Hui:2009kc}:
\be\label{ansatz}
\mathcal X=\mathcal X_0+\mathcal X_1(r)\,,\, \mathcal X_0(x^i)\simeq \mathcal X_0(0)+\partial_i\mathcal X_0(0)x^i\,,
\ee
where $\mathcal X_1$ represent the fields due to the object itself while $\mathcal X_0$ amount to some large scale background fields that vary mildly on the scale of the sphere. We do not specify the origin of these background fields, but use the freedom to add such linear gradient fields to any other solution to equations consisting of second order spatial derivatives. This is the case in the quasi-static subhorizon limit. For performing the integrals in \eqref{dotPi} we use the following useful relationships for $\{\mathcal X,\mathcal Y\}$ satisfying the ansatz \eqref{ansatz}:\footnote{We have also made use of the identities $\partial_i \mathcal X_1(r)=n_i\mathcal X_1'$, $\partial_i\partial_j \mathcal X_1(r)=n_i n_j\mathcal  X_1''+(\delta_{ij}-n_in_j)\frac{\mathcal X_1'}{r}$, $\oint dA\cdot n_j=0$ and $\oint dA\cdot n_i n_j = \frac{4\pi}{3}r^2\delta_{ij}$.}
\begin{align}
&\oint dS_i\,\partial^i\mathcal X\,\partial_j \mathcal Y=\frac{4\pi}{3}r^2\partial_j \mathcal X_0Y'_1+4\pi r^2\partial_j\mathcal Y_0\mathcal X'_1\,,\nonumber\\
&\oint dS_i\,\mathcal X(\delta_{ij}\Delta\mathcal Y-\partial_i\partial_j\mathcal Y)=\frac{8\pi}{3}r^2\partial_j\mathcal X_0\mathcal Y'_1\,,\nonumber\\
&\oint dS_j\,\partial_k\mathcal X\,\partial^k\mathcal X=\frac{8\pi}{3}r^2\partial_j\mathcal X_0\mathcal X'_1\,.
\end{align}  
By performing this split in the last term of \eqref{dotPi} as $\Phi=\Phi_0+\Phi_1(r)$, we see that the only surviving term is the one coming from $\partial_i\Phi_0$. Assuming that inside our volume of integration the mass of the object is larger than the total mass of the background matter fields characterized by $\rho_{\text m}, p_{\text m}$, allows us to neglect the last term in \eqref{dotPi}.

\subsection{Test particles}
It is instructive to start by looking at the case of test particles, even though the result is trivial---their inertial and gravitational masses are equal and they move on geodesics.
The advantage of test particles is that the perturbative expansion is a valid approximation everywhere. In particular, the integral \eqref{nonlinear} can be evaluated at quadratic order in perturbations by making use of the following useful relationship, derived from~\eqref{eqpi}:
\be\label{quadratic}
H\tau^{(2)i}\,_i+\mathcal E_{(2,\pi)}=\partial_iT^{(2,\pi)i}\,_0\,.
\ee
The above relationship is derived in subhorizon approximation, \emph{i.e.} in the leading order of derivatives. For the derivative of the ADM mass we thus obtain:
\be
\dot M=-\M^2\oint dS_ja^3G^{(2)}\,^j\,_0=0\,.
\ee
Similarly, for the geodesic equation \eqref{geodesic} we find:
\begin{align}\label{geodesic_new}
\ddot X^i+&2H\dot X^i=\\
&\frac{1}{a^2M}\,\partial_0\left[P_i-\int d^3x\,a^3\,T^{(2,\pi)0}\,_i\right]\,.\nonumber
\end{align}
We stress that this equation was derived by only assuming the validity of the perturbative expansion without ever referring to the exact form of the action. We have then used the scalar field equation of motion \eqref{eqpi} together with the conservation laws of the pseudo stress tensor in \eqref{cons0} and \eqref{consi}. The equation \eqref{geodesic_new} is, however, important---it states that in the presence of a scalar field the force that is felt by a test particle moving in the gravitational field is not given by the time derivative of momentum as defined in \eqref{defPi}. Instead, the contribution of the scalar field has to be subtracted. Physically this shows that the gravitational force felt by a test particle is not altered by the presence of the scalar. In general, however, the nonlinear effects modify the geodesic equation as in \eqref{geodesic}.

As was said above, for the test particle, all the nonlinear quantities can be evaluated at quadratic order. By explicit calculation for the action \eqref{act_cov} we then find
\be\label{dotPiv2}
\begin{split}
\dot P_i-\partial_0\int d^3x\,&a^3T^{(\pi,2)}\,^0_i=\\
&\partial_i\Phi_0\int d^3x\,a^3\left(\tau^0\,_0-\frac{1}{3}\tau^i\,_i\right)\\
+&\partial_i\pi_0\int d^3x\, a^3\mathcal E_{\rm{NL},\pi}\,.
\end{split}\,
\ee
We note that the right-hand side of this expression is completely nonlinear and we have not assumed validity of the perturbative expansion. We have only used the explicit form of the quadratic stress-energy tensor, together with the equations of motion \eqref{eq00lin}--\eqref{eqpilin} written through the nonlinear quantities $\tau^\mu\,_\nu$. 
For test particles, the nonlinear term on the last line can be dropped, and for a non-relativistic source $\tau^0\,_0\gg\tau^i\,_i$. We thus get that the only relevant contribution on the right-hand side of the above equation is $-M\partial_i\Phi_0$, where $M$ stands for the ADM mass defined in~\eqref{defM}. Inserting this in \eqref{geodesic_new} we recover the correct geodesic equation
\be\label{geodesic0}
\ddot X^i+2\frac{\dot a}{a}\dot X^i=-\frac{1}{a^2}\partial_i\Phi_0\,.
\ee
We present the detailed derivation of Eq.~\eqref{dotPiv2} together with the expressions for the quadratic components of the stress-energy tensor $T^\pi_{\mu\nu}$ in Appendix~\ref{sec:emt}.

\subsection{Extended objects}\label{sec:extended}
In this section we shall extend the derivation of the gravitational force felt by a test particle to the case of extended objects. In this case we assume that the self-gravitational energy is still negligible, however screening effects close to the object need to be taken into account. Technically this means that far from the object, \emph{i.e.} outside its Vainshtein radius, the fields $\Phi,\Psi,\pi$ and their derivatives are still in the perturbative regime. Inside the Vainshtein radius, however, while the fields themselves can still be treated perturbatively, the higher derivative terms $\partial^2\mathcal X$ for $\mathcal X\in\{\Phi,\Psi,\pi\}$ become dominant and need to be taken into account. 

Let us start by considering the time derivative of the center of mass coordinate, given in \eqref{dotXi}. In particular, let us discuss the problematic integral $\mathcal I$ defined in~\eqref{nonlinear}. The integrand here is a fully nonlinear expression. In the case of test particles the perturbative expansion was valid everywhere and we could use the quadratic expression \eqref{quadratic} in order to evaluate this integral. The situation is different for sizeable objects---in the close vicinity to the object, \emph{i.e.} inside its Vainshtein radius $r_V$, the nonlinear derivative interactions of the scalar field dominate. However, as was discussed in great detail in Sec.~\ref{extended}, in theories described by the action~\eqref{act_cov}, the leading Vainshtein interactions arise at the cubic order. These result in the following non-trivial contributions in the integrand of \eqref{nonlinear}:
\begin{equation*}
\mathcal I^{\text{lead}}=\int d^3x\,x^i\,a^3\left[H\tau^{(\pi,2)j}\,_j+\mathcal E_{2,\pi}+H\tau^{(\pi,3)j}\,_j\right]\,.
\end{equation*}
All higher order contributions will be subleading. We further note that the derivative structure of these leading terms is $\tau^{(\pi,2)j}\,_j\sim(\partial\pi)^2$, $\mathcal E_{2,\pi}\sim (\partial^2\pi)^2$ and $\tau^{(\pi,3)j}\,_j\sim(\partial\pi)^2\partial^2\pi$. It is known that inside the Vainshtein region $\partial^2\pi\gg(\partial\pi)^2$. Hence, the cubic term above can be neglected and the integrand of \eqref{nonlinear} contains contributions that cannot be neglected for sizable objects with negligible self-gravity only in terms up to quadratic order in perturbations. As for the test particle, it can be evaluated by using the relationship \eqref{quadratic}, giving
\be
\mathcal I^{(2)}=\int d^3x\,x^i\,a^3\partial_jT^{(\pi,2)j}\,_0=-\int d^3x\,a^3\,T^{(\pi,2)i}\,_0\,,
\ee
where we have dropped a surface integral over energy flux in the last equality. 

In order to derive the geodesic equation we substitute the above result in \eqref{geodesic}, leading to \eqref{geodesic_new}. In distinction from the case of the test particles, however, we have to take into account also the cubic contributions to the time derivative of the momentum $P_i$. Thus, at the leading order in the subhorizon limit instead of Eq.~\eqref{dotPiv2} we get:
\be\label{dotPiv3}
\begin{split}
\dot P_i-&\partial_0\int d^3x\,a^3T^{(\pi,2)}\,^0_i=\\
&\partial_i\Phi_0\int d^3x\,a^3\left(\tau^0\,_0-\frac{1}{3}\tau^i\,_i\right)\\
+&\partial_i\pi_0\int d^3x\, a^3\mathcal E_{\rm{2},\pi}-\oint dS_j\,a^3T^{(\pi,3) j}\,_i\,.
\end{split}\,
\ee
At this point the fact that we are dealing with an extended object again becomes relevant. It can still be taken to be non-relativistic, so that the terms on the second line can be expressed in terms of the ADM mass as in the case of the test particles, leading to $-M\partial_i\Phi_0$. However, the cubic terms on the second line are slightly more problematic. In particular, there is a remaining volume integral which we cannot evaluate without knowing the full nonlinear field solutions also within the Vainshtein regime. A resolution to this comes from the approximate shift invariance of the theory. This allows one to write the equation of motion for the scalar field in the subhorizon limit in the form of a spatial divergence as in \eqref{divergence}. In fact, it takes the form of a \emph{twofold} spatial divergence:\footnote{This observation holds also in the case when the quartic Vainshtein interactions are present in the action. See Appendix~\ref{cubic} for the explicit form.}
\begin{align}\label{eq_quadr}
\mathcal E^{\text{lead}}_{2,\pi}&=-\frac{m_3^3}{a^4}\partial_i\partial_j\left[\partial_i\pi\partial_j\pi-\delta_{ij}(\partial\pi)^2\right]\\
&=\partial_iT^{(2,\pi)i}\,_0\,.\nonumber
\end{align}
The last equality follows from the general relation~\eqref{quadratic} (given that in the quasi-static subhorizon limit $\tau^{(2)i}\,_i\ll\mathcal E^{\text{lead}}_{2,\pi}$), but can also be seen to hold from the explicit form \eqref{defTi0}. 

There are then two possible ways to deal with the cubic terms on the last line of~\eqref{dotPiv3}. The simplest way of evaluating them is to use the explicit form of the scalar equation of motion to rewrite the cubic terms as surface integrals
\be\label{surf_int}
\oint dS_j\,a^3\left(\partial_i\pi_0\, T^{(2,\pi)j}\,_0-T^{(\pi,3) j}\,_i\right)\,.
\ee
We can then choose the integration surface outside the Vainshtein radius of the object, where the field perturbations are in the linear regime and the arising contributions from the integrals are negligible. By substituting the remaining terms of \eqref{dotPiv3} in \eqref{geodesic_new} we then arrive at the geodesic equation
\be\label{geodesic_extended}
\begin{split}
\ddot X^i+&2\frac{\dot a}{a}\dot X^i=-\frac{1}{a^2}\partial_i\Phi_0\\
&+\frac{1}{a^2MH}\partial_i(H\pi_0)\int d^3x\, a^3\mathcal E^{\rm{sublead}}_{\rm{NL},\pi}\,,
\end{split}
\ee
where we have added the factors of $H$ in the last term, so that both $\partial_i\Phi_0$ and $\partial_i(H\pi_0)$ have the same dimensionality.  

One might be rightly concerned that taking the integration surface so far away from the object the background-object split is not valid anymore. Indeed, for any ordinary object its Vainshtein radius is much larger than its size. Moreover, it overlaps with the Vainshtein radii of other objects, even when the physical distance between the actual objects is large. Thus it is certainly not reasonable to assume that the background fields can be approximated as linear divergence fields over such large scales. In \cite{Hui:2009kc} it was, however, argued that it is sufficient if one can approximate the background field as a linear gradient field in a neighbourhood close to the object.\footnote{For limitations of this assumption in the deeply non-linear regime when the Vainshtein regions of the test object and the object sourcing the background field overlap, see~\cite{Hiramatsu:2012xj}.}  Due to the shift symmetry of the derivative, $\partial_\mu\pi\to\partial_\mu\pi+c_\mu$, one can then extrapolate the linear background field close to the object up to distances exceeding the Vainshtein radius. This argument cannot be straightforwardly applied on FRW backgrounds, even though one would expect the argument to hold in the subhorizon limit. 

Let us prove that the geodesic equation \eqref{geodesic_extended} is in fact the correct result for the equation of motion of an extended object moving in a linear background field. For this we shall evaluate the integral \eqref{surf_int} exactly, on a surface close to the object where the background fields can safely be assumed to be linear gradient fields. By performing the background-object split in the expression \eqref{cubic_emt} of the cubic stress-energy tensor components, we obtain for its surface integral:
\be
\oint dS_j\,T^{(\pi,3)j}\,_i=-\frac{2m_3^3}{a^4}\partial_i\pi_0\frac{(\pi_1')^2}{r}\cdot 4\pi r^2\,,
\ee
where we have used that $\oint dA\cdot n_in_jn_k=0$. Similarly, for the other integral in \eqref{surf_int} we find:
\be
\oint dS_j \,T^{(2,\pi)j}\,_0=-\frac{2m_3^3}{a^4}\frac{(\pi_1')^2}{r}\cdot 4\pi r^2\,.
\ee
Combining the two we see that the total cubic contribution \eqref{surf_int} vanishes without specifying the radius of the integration surface $r$. Hence, the geodesic equation \eqref{geodesic_extended} is indeed our final result. We thus conclude that the equivalence principle holds for screened extended objects, up to some violations due to the subleading terms in the equation for $\pi$. These arise from the terms that do not come in the form of a spatial divergence and will be evaluated in the next section.

At last, let us note that starting from Eq.~\eqref{geodesic} we have dropped the contributions of the order $\mathcal O(H\dot M/M)$ to the geodesic equation. Given the expression \eqref{dotM} one can argue,  along the same lines as above for the integral $\mathcal I$, that the ADM mass is also conserved for extended objects. 

\subsection{Estimate for violation}\label{sec:violations}
In this section we would like to estimate the order of magnitude of the violations of the equivalence principle for extended objects. According to our computations, the only modification to the geodesic equation is of the form \eqref{geodesic_extended}. As we emphasized already, the Vainshtein mechanism takes place due to the cubic order interactions in the EFT action. Hence, the relevant nonlinear contributions to the equations of motion truncate at second order and are given by $\mathcal E^{\text{lead}}_{2,\pi}$ in \eqref{eq_quadr}. We are interested in evaluating the contributions from the next-to-leading terms appearing at quadratic level. These are, for instance,
\be\label{nlo}
\mathcal E^{\rm sublead}_{2,\pi}=-\frac{m_3^3}{2a^2}H^2\partial_i\pi\partial_i\pi\,.
\ee
By performing the object-background split \eqref{ansatz} we find that the volume integral over these terms gives
\be\label{correction}
\begin{split}
\int d^3x\,\mathcal E^{\rm sublead}_{2,\pi}&=m_3^3 H^2\int d^3x\,\left[\left(\partial_i\pi_0\right)^2+\left(n_i\pi_1'\right)^2\right]\\
&=4\pi m_3^3H^2\int dr\,r^2\,\left(\pi_1'\right)^2\,,
\end{split}
\ee
where in the last equality we have dropped the contributions coming from the background since the integral should be dominated by the contributions from the object itself. We note that we are integrating over a sphere of radius $r$ that is enclosing the object, but is large enough so that the field $\pi_1$ is determined by its asymptotic monopole solution. On the other hand the sphere should be small enough that the external gravitational fields can be approximated as linear gradient fields. 

\subsubsection{Linear solution}
Let us first evaluate \eqref{correction} for $\pi_1(r)$ given by the asymptotic solution from the linear equations of motion, obtained by combining eqns. \eqref{eq00lin}-\eqref{eqpilin}:
\be
\left[-\frac{2\M^2}{m_3^3}c-2\M^2H+m_3^3\right]\pi_1'=\frac{M}{a}\frac{1}{4\pi r^2}\,.
\ee
For a generic modification of gravity, all the terms inside the square brackets are of the same order:
\be\label{order}
\frac{\M^2}{m_3^3}c\sim\M^2H\sim m_3^3\,.
\ee 
We can thus write an order of magnitude estimate
\be\label{pi_lin}
\pi_1'\sim\frac{M}{m_3^3}\frac{1}{r^2}\,.
\ee

\subsubsection{Vainshtein solution}
Next, let us estimate the Vainshtein scale for the theory at hand. The only equation in which the Vainshtein nonlinearities play a significant role is the scalar field equation of motion. In the subhorizon limit this reads:
\be\label{eom_vainsh}
\begin{split}
\frac{1}{a^2}&\left[2c\Delta\pi+m_3^3\left(H\Delta\pi+\Delta\Phi\right)\right]\\
&=\frac{m_3^3}{a^4}\partial_i\partial_j\left[\partial_i\pi\partial_j\pi-\delta_{ij}(\partial\pi)^2\right]\,.
\end{split}
\ee
We shall fix the Hubble parameter to its present value and set $a(t)=a_0=1$ in this section; this will give correct order of magnitude estimates. 
The Vainshtein scale is defined as the scale at which the nonlinear terms on the right-hand side become of the same order as the linear terms on the left-hand side of this equation. We can estimate the scale by schematically comparing
\be
m_3^3 \partial^2(\partial\pi)^2\sim c\,\partial^2\pi\sim m_3^3\partial^2\Phi \,,
\ee
where in the spherically symmetric case we can further replace $\partial\to r^{-1}$. 
From the first 'equality' in the above relation we then obtain that at the Vainshtein radius $\pi\sim c\,r_V^2/m_3^3$. By equating this with the linear solution \eqref{pi_lin}  and using $c\sim m_3^6/\M^2$ from \eqref{order}, we get for the Vainshtein scale
\be
r_V=\left(\frac{M\M^2}{m_3^6}\right)^{1/3}\,.
\ee
We can then find the spherically symmetric solution for $\pi$ deep inside the Vainshtein radius, when the nonlinear terms in \eqref{eom_vainsh} dominate. Simplifying the above expression as $r^{-2}\Phi\sim r^4\pi^2$ and taking the linear solution for $\Phi\sim M/\M^2/r_V$, we obtain the screened solution
\be\label{inside}
\pi_{\rm screen}\sim \sqrt{\frac{Mr}{\M^2}}\,,\qquad r\ll r_V\,.
\ee

\subsubsection{The corrections}
Now we are able to evaluate the correction \eqref{correction} by splitting the integral of the subleading terms in~\eqref{correction} as 
\be\label{int_split}
\begin{split}
\int dr\,r^2\,\left(\pi_1'\right)^2&=\int_0^{r_V} dr\,r^2\,\left(\partial_r\pi_{\rm screen}\right)^2\\
&+\int_{r_V}^r dr'\,r'^2\,\left(\partial_{r'}\pi_1\right)^2\,.
\end{split}
\ee
Given the functional forms of the solutions \eqref{inside} and \eqref{pi_lin} we see that both integrals in \eqref{int_split} give their main contribution at $r=r_V$ and are of the same order of magnitude:
\be
\frac{\int^{r_V} dr\,r^2\,\left(\partial_r\pi_{\rm screen}\right)^2}{\int^{r_V} dr\,r^2\,\left(\partial_r\pi_{1}\right)^2}\sim\frac{\frac{r_V^2M}{\M^2}}{\frac{M^2}{m_3^6 r_V}}=\frac{r_V^3m_3^6}{M\M^2}=\mathcal O(1)\,.
\ee
To evaluate the contribution of the subleading terms  to the geodesic equation \eqref{geodesic_extended} we can therefore approximate it by evaluating one of the integrals at the Vainshtein radius. After recasting the result in a more meaningful form we find
\be\label{corr1}
\begin{split}
\frac{H^{-1}}{M}\int d^3x\,\mathcal E^{\rm sublead}_{\rm 2,\pi}\sim\frac{r_g}{r_V}\,,
\end{split}
\ee
where $r_g=M/\M^2$ is the Schwarzschild radius of the object and we have used $H\M^2\sim m_3^3$.\footnote{Let us remark that assuming $H\M^2\sim m_3^3$ is equivalent to relating the Vainshtein radius, Schwarzschild radius and the Hubble scale as $r_VH\sim (r_g/r_V)^{1/2}$. It is, in fact, the case in massive gravity theory where $r_V^3\sim M/(\M^2 m^2)$ and the graviton mass is taken to be $m\sim H$. } For an object of mass $M_0$, the ratio $r_g/r_V$ equals to
\be
\frac{r_g}{r_V}\sim\left(\frac{HM_0}{\M^2}\right)^{2/3}\ll1\,.
\ee
For any astrophysical object this ratio is very small. As an example, for the Sun this ratio is $\sim10^{-15}$. In fact, the mass of the object should coincide with the mass of the whole universe for this ratio to be of order one. 

As we have emphasized already, for ordinary objects the Vainshtein radius is much larger than their size. Therefore, for the background-object split to be valid, the integration volume in \eqref{int_split} should only extend up to $r\ll r_V$. In this case we find an additional suppression factor to \eqref{corr1}:
\be
\begin{split}
\frac{H^{-1}}{M}\int d^3x\,\mathcal E^{\rm sublead}_{\rm 2,\pi}\sim\left(\frac{r}{r_V}\right)^2\frac{r_g}{r_V}\,,
\end{split}
\ee
leading to a modified geodesic equation of the form:\footnote{Without the assumption $m_3^3\sim H\M^2$ the scales $r_V,r_g, H^{-1}$ decouple and the corrections can be recast as $\left(\frac{r}{r_V}\right)^{3/2}\left(\frac{r}{H^{-1}}\right)\left(\frac{r_g}{r}\right)^{1/2}$. We recognize the power $3/2$ as the power of the Vainshtein corrections to the gravitational potential in the DGP theory \cite{Dvali:2002vf}. }
\be\label{res_viol}
\ddot X^i=-\partial_i\Phi_0+\partial_i(H\pi_0)\left(\frac{r}{r_V}\right)^2\frac{r_g}{r_V}\,,\quad r\ll r_V\,.
\ee
Again, for the example of the Sun, the ratio $R_{\odot}/r_V\sim 10^{-10}$. This ratio is somewhat larger for galaxies. Combining the two suppression factors we thus conclude that the violation of the equivalence principle is very tiny. In principle, if one could measure deviations from the geodesic motion (if any) up to this level of precision one could use (a more accurate version of) these estimates of the deviations from the equivalence principle to put constraints on the EFTs of dark energy described by the action \eqref{act_cov}.

\section{Conclusions and Discussion}\label{conclusions}
Without doubt the equivalence principle is one of the cornerstone predictions of General Relativity. In this work we have addressed the question of its violation on cosmological backgrounds in scalar-tensor theories of modified gravity. We find that an exactly de Sitter spacetime stands out as a particular gravitational background where both the weak and the strong equivalence principles are obeyed. Any violations of the equivalence principle can therefore be parametrized in terms of the time derivatives of the Hubble rate. For slowly changing Hubble rate, which is the situation in the present day universe, these are small and can thus be treated perturbatively. This provides new ways for characterizing the violations of the equivalence principle that comes particularly timely given the new gravity wave observations opening up the possibility of testing gravity in the strong-field regime. 

We have also investigated the motion of weakly gravitating objects on FRW spacetimes with arbitrary time dependence of the Hubble rate. We show that the weak equivalence principle is obeyed for test particles and extended objects moving in some external gravitational potential on these backgrounds. 
This result complements the findings of \cite{Barausse:2015wia,Lehebel:2017fag,Barausse:2017gip}  that in shift symmetric Horndeski theories in flat space the star solutions bear zero scalar charge and thus have no-hair. Consequently, the gravitational and inertial masses of compact stars are equal, \emph{i.e.} they obey the equivalence principle, and the dipolar gravitational wave emission vanishes \cite{Barausse:2015wia}. However, in distinction from \cite{Barausse:2015wia,Lehebel:2017fag,Barausse:2017gip} we do not claim that the scalar field profile is vanishing. Instead we show that the scalar hair is secondary and is determined entirely in terms of the ADM mass of the object. As a result the extended objects move on geodesics, as in General Relativity. 

We note, however, that the proof given in \cite{Barausse:2015wia} relies on the assumption that the Vainshtein radius of the star is much less than the gravitational wavelength, \emph{i.e.} that $r_V\ll\lambda_{\rm GW}$. This is in general not true. For compact stars of masses close to the solar mass, the Vainshtein radius is $r_V\sim 10^{15}$~km while the wavelength of the gravity waves is $\lambda_{\rm GW}\sim 10^4$~km. In this work we have shown that compact objects move on geodesics also on distances less than its Vainshtein radius. The only condition we use is that in the close vicinity of the object the total external gravitational field can be treated as a linear gradient field. For a binary system of two neutron stars this would mean that our conclusion is valid for distances much less than the binary separation, while one is still allowed to treat the two binaries in isolation \cite{Barausse:2017gip}. 

Finally, we have also presented an example of how the corrections to the object's geodesic equation can be computed. The result given in~\eqref{res_viol} shows the scaling of the violations of the equivalence principle at a given distance $r$ from the extended object with respect to the various length scales, $r_V, r_g, H^{-1}$. This includes all the assumptions made in our derivation: $(i)$ subhorizon limit, $r\ll H^{-1}$; $(ii)$ weak gravitation, $r\gg r_g$; $(iii)$ distances less than the Vainshtein radius, $r\ll r_V$. By focusing on a specific scalar-tensor model one can derive this result to an arbitrary level of precision that can then be constrained by the gravitational wave measurements. We leave this for future work. 

\section*{Acknowledgements}
We are particularly grateful to 
Paolo Creminelli, J\'er\^ome Gleyzes 
and Filippo Vernizzi 
for many useful discussions and for collaboration in the early stages of this project. We thank Lam Hui and Jeremy Sakstein for insightful discussions. We further thank Claudia de Rham, Alberto Nicolis and Andrew Tolley for many useful comments on the manuscript. LA is supported by European Union's Horizon 2020 Research Council grant 724659 MassiveCosmo ERC-2016-COG and by the European Research Council under the European Union's Seventh Framework Programme (FP7/2007-2013), 
ERC Grant agreement ADG 339140.


\appendix

\section{Current components}\label{app:current}
To obtain the covariant action we do the following replacements in the unitary gauge action \eqref{act_unit}:
\begin{align}\label{replace1}
&\delta g^{00}\to 1+\frac{(\partial\phi)^2}{\M^4}\,,\\\label{replace2}
&\delta K\to -3H\left(\frac{\phi}{\M^2}\right)-\frac{3}{2\M^2}\Box\phi\\&-\frac{1}{2\M^6}\left[(\partial\phi)^2\Box\phi+\partial^\mu\phi\partial_\mu(\partial\phi)^2\right]\,.\nonumber
\end{align}
We note that the expression for the extrinsic curvature coincides with the actual expression, $K \equiv \nabla_\mu n^\mu$, with $n_\mu\equiv -\partial_\mu\phi/\sqrt{-(\partial\phi)^2}$ only up to second order in perturbations. As a result, the actions~\eqref{act_unit} and \eqref{act_cov} coincide only up to cubic order. 
The equation of motion for the scalar field $\phi$ can then be derived from the covariant action \eqref{act_cov} using the equations \eqref{eom_current} and \eqref{current}. We repeat them here for convenience:
\begin{align}\label{eom_app}
&\frac{\partial\mathcal L}{\partial\phi}-\nabla_\mu J^\mu=0\, ,\\
&J^\mu=\frac{\partial\mathcal L}{\partial(\nabla_\mu\phi)}-\nabla_\nu\left(\frac{\partial\mathcal L}{\partial(\nabla_\mu\nabla_\nu\phi)}\right)\,.
\end{align}
The relevant expressions are:
\begin{align}
&\frac{\partial\mathcal L}{\partial(\nabla_\mu\phi)}=\\
&-2c\frac{\nabla^\mu\phi}{\M^4}+2m_2^4\frac{\nabla^\mu\phi}{\M^4}\delta g^{00}-m_3^3\frac{\nabla^\mu\phi}{\M^4}\delta K\nonumber\\
&+\frac{m_3^3}{2\M^6}\delta g^{00}\left(\nabla^\mu\phi\Box\phi+2\nabla_\nu\phi\nabla^\nu\nabla^\mu\phi\right)\nonumber
\end{align}
and
\begin{align}
&\frac{\partial\mathcal L}{\partial(\nabla_\mu\nabla_\nu\phi)}=-\frac{m_3^3}{2}\delta g^{00}\left(-\frac{3}{2\M^2}g^{\mu\nu}\right.\nonumber\\
&\quad\left.-\frac{1}{2\M^6}\left[(\partial\phi)^2g^{\mu\nu}+2\nabla^\mu\phi\nabla^\nu\phi\right]\right)\,.
\end{align}
The part of the equation of motion \eqref{eom_current} that cannot be expressed as a conservation law is contained in
\begin{align}
\label{non_cons}
\frac{\partial \mathcal L}{\partial\phi}&=-\partial_\phi\Lambda-\frac{(\partial\phi)^2}{\M^4}\, \partial_\phi c+\frac{3}{2}m_3^3\,\delta g^{00}\,\partial_\phi H\,\nonumber\\
&+\frac{1}{2}\partial_\phi m_2^4\,(\delta g^{00})^2-\frac{1}{2}\partial_\phi m_3^3\,\delta g^{00}\delta K\,.
\end{align}
In all the above expressions we keep in mind that the parameters $\Lambda, c, H, m_2, m_3$ all depend on the scalar field as $\Lambda=\Lambda\left(\frac{\phi}{\M^2}\right)$ etc. For the sake of readability we have also kept using $\delta g^{00}$ and $\delta K$ instead of expressing them through \eqref{replace1} and \eqref{replace2}. The equation of motion \eqref{eom_app} then reproduces the equation of motion for $\pi$ as it would be obtained from the action \eqref{act_unit} after the St\"uckelberg trick $t\to t + \pi(t,x^i)$ up to second order in perturbations. By doing the expansion in perturbations one can see that in the quasi-static subhorizon limit (\textit{i.e.} when $\partial_t^2\pi\ll\partial_i^2\pi$ and $H^2\pi\ll \partial_i^2\pi$) the contributions coming from the terms \eqref{non_cons} due to the explicit violation of the shift symmetry are subleading. 

\section{Derivation for test particles}\label{sec:emt}
\subsection{Quadratic and cubic tensor components}
Here we give the relevant expressions needed for the derivation of the geodesic equation for a test particle in the subhorizon limit. The quadratic $ij$ components of the stress-energy tensor of $\pi$ read
\be\label{t2noexp}
\begin{split}
T^{(\pi,2)}\,^i_j&=\frac{\M^2}{a^2}\bigg\{\frac{2c}{\M^2 }\partial_i\pi\partial_j\pi+2\frac{m_3^3}{\M^2}\,\partial_{(i}\pi\partial_{j)}(\Phi-\dot\pi)\\
&+\delta^i_j\bigg[\frac{m_3^3}{\M^2}\left(\frac{H}2(\partial_k\pi)^2+2\partial_k\pi\partial_k\dot\pi-\partial_k\pi\partial_k\Phi\right)\\
&-\frac{c}{\M^2}(\partial_k\pi)^2+\frac{(m_3^3)^{\hbox{$ \cdot$}}}{2\M^2 }\partial_k\pi\partial_k\pi\bigg]\bigg\}\,.
\end{split}
\ee
The quadratic contribution to the Einstein's tensor in the Newtonian gauge \eqref{gauge} is given by:
\begin{align*}
a^2\,&G^{(2)}\,_i\,^j=
\partial_i\Psi\partial_j\Psi-\partial_i\Phi\partial_j\Phi-2\partial_{(i}\Phi\partial_{j)}\Psi
\nonumber\\
&+2\Psi\partial_i\partial_j(\Psi-\Phi)+\left[(\partial_k\Phi)^2-2\Psi\Delta(\Psi-\Phi)\right]\delta_{ij}
\,.
\end{align*}
The leading order $0i$ components are in turn
\begin{align}
\label{defTi0}
T^{(\pi,2)}\,^0\,_i=&\frac{m_3^3}{a^2}\left[\partial_i\pi\Delta\pi-\partial_j\pi\partial_j\partial_i\pi\right]\nonumber\\
=&\frac{m_3^3}{a^2}\partial_j\left[\partial_i\pi\partial_j\pi-\delta_{ij}(\partial\pi)^2\right]\,,
\end{align}
while $G^{(2)}\,^i_0$ is subdominant. 

We shall also need the cubic $ij$ components of the stress tensor that in the subhorizon limit read:
\be\label{cubic_emt}
\begin{split}
T^{(\pi,3)i}\,_{j}=-\frac{m_3^3}{a^4}\left[\delta_{ij}\partial_k\pi\partial_l\pi\partial_k\partial_l\pi+\Delta\pi\partial_i\pi\partial_j\pi\right.\\
\left.-\partial_i\pi\partial_k\pi\partial_j\partial_k\pi-\partial_j\pi\partial_k\pi\partial_i\partial_k\pi\right]\,.
\end{split}
\ee

\subsection{Gravitational force}
In order to find the gravitational force exerted on the moving object, according to \eqref{geodesic_new} we need to first compute the time derivative of the momentum $P_i$ defined in \eqref{defPi}:
\begin{align}
\dot P_i=a^3\oint dS_j\,\left(-T^{(\pi,2)}\,_i\,^j+\M^2G^{(2)}\,_i\,^j\right)\,.
\end{align}
Using the background-object split \eqref{ansatz} we find for the second term:
\be\label{gr_contr}
a^2\oint dS_j\,G^{(2)}\,_i\,^j=-\frac{8\pi}{3}r^2\partial_i\Phi_0\,\left(\Phi_1'+2\Psi_1'\right)\,.
\ee
The surface integral of the first term combined with the time derivative of the $0i$ component of the stress tensor, as in the geodesic equation \eqref{geodesic_new}, in turn becomes 
\begin{equation*}
\begin{split}
-a^3\oint dS_j\,T^{(\pi,2)}\,^i_j&-\partial_0\int d^3x\,a^3T^{(\pi,2)}\,^0_i=\\
-4\pi r^2a\partial_i\Phi_0&m_3^3\pi_1' \\
-4\pi r^2a\partial_i\pi_0&\bigg(m_3^3\Phi_1'+(2c+(m_3^3)^{\hbox{$ \cdot$}}+Hm_3^3)\pi_1'\bigg)\,,
\end{split}\,
\end{equation*}
where we have performed the background--object split according to the ansatz \eqref{ansatz}.  After integrating both sides of the eqs.~\eqref{eq00lin}--\eqref{eqpilin} over a volume enclosing the object, we express the fields $\pi', \Phi',\Psi'$ through the fully nonlinear solutions to arrive to:
\be
\begin{split}
-a^3\oint dS_j\,&T^{(\pi,2)}\,^i_j-\partial_0\int d^3x\,a^3T^{(\pi,2)}\,^0_i=\\
+&\frac{8\pi}{3} r^2a\M^2\partial_i\Phi_0(2\Psi_1'+\Phi_1')\\
+&\partial_i\Phi_0\int d^3x\,a^3\left(\tau^0\,_0-\frac{1}{3}\tau^i\,_i\right)\\
+&\partial_i\pi_0\int d^3x\, a^3\mathcal E_{\rm{NL},\pi}\,.
\end{split}\,
\ee
We stress that this relationship is completely nonlinear and we have not assumed that the perturbative expansion is valid. We have simply used the form of the quadratic stress-energy tensor, together with the equations of motion \eqref{eq00lin}--\eqref{eqpilin}, written through the nonlinear quantities $\tau^\mu\,_\nu$. 
For test particles, the nonlinear term on the last line can be dropped, and for a non-relativistic source $\tau^0\,_0\gg\tau^i\,_i$. In combination with the contribution \eqref{gr_contr} arising from the Einstein tensor we thus get
\be
\dot P_i-\partial_0\int d^3x\,a^3T^{(\pi,2)}\,^0_i=-M\partial_i\Phi_0\,,
\ee
where $M$ stands for the ADM mass defined in~\eqref{defM}. Inserting this in \eqref{geodesic_new} we recover the correct geodesic equation
\eqref{geodesic0}.
\bigskip

\section{Current conservation}\label{sec:current2}
\subsection{Linear evolution}
The full  expressions for the linearized components of the current are (also including the time dependence of $m_2$ and $m_3$):
\begin{align}
J^0&=-\frac{1}{\M^2a^2}\left[m_3^3\Delta\pi-3a^2H\rho_m\pi+4m_2^4a^2\Phi\right.\nonumber\\
&+3m_3^3a^2H\Phi-2a^2\rho_m\Phi+3m_3^3a^2\dot H\pi\nonumber\\
&-4\M^2a^2\dot H\Phi+2\M^2a^2\ddot H\pi-4m_2^4a^2\dot\pi\nonumber\\
&\left.+a^2\rho_m\dot\pi+2\M^2a^2\dot H\dot\pi+3m_3^3a^2\dot\Psi\right]\,,
\end{align}
while the spatial current reads:
\be
J^i=\frac{1}{\M^2a^2}\partial_i\left(-2c\pi-m_3^3\Phi+m_3^3\dot\pi\right)\,.
\ee
We also note that the non-conserved part of the equation of motion is indeed subleading in the quasi-static subhorizon limit:
\begin{align}
-\M^2\frac{\partial\mathcal L}{\partial\phi}=&-\Phi\left(3m_3^3\dot H+\dot\rho_m+2\M^2\ddot H\right)\nonumber\\
&+2\M^2\pi\left(3\dot H^2+3H\ddot H+\dddot H\right)\nonumber\\
&+\left(3m_3^3\dot H+\dot\rho_m+2\M^2\ddot H\right)\dot\pi\,.
\end{align}
The leading terms in the linear equation of motion as derived from \eqref{current} are thus given by the divergence of the current $J^\mu$ as:
\be
\mathcal E_{{\rm L},\pi}=\frac{1}{a^2}\left[2c\Delta\pi+m_3^3\left(H\Delta\pi+\Delta\Phi\right)\right]\,.
\ee
This coincides with the linear part of the equation \eqref{eqpilin}. Importantly, we note that the $m_3^3H\Delta\pi$ term comes from the $J^0$ component. This is very different from the case of static backgrounds (see, \emph{e.g.}~\cite{Hui:2012qt}).

\subsection{Quadratic equation}
Let us first comment on the non-conserved part of the current conservation equation, $\partial\mathcal L/\partial\phi$, given in Eq.~\eqref{non_cons}. First, it is clear that at any order in perturbations, the terms on the first line in \eqref{non_cons} will contain at most two derivatives of the form $(\partial\pi)^2$. In the same time we know that there will be terms with more derivatives arising in the full equation of motion for the scalar. Second, the terms on the second line, although in general of the same order as $\nabla_\mu J^\mu$, will not contain the dominant Vainshtein interactions arising at the quadratic order. 
Hence, it is safe to neglect the $\partial\mathcal L/\partial\phi$ term in the quasi-static subhorizon limit, so that
\be\label{quadr_eom}
\mathcal E_{2,\pi}=\left(-\M^2\nabla_\mu J^\mu\right)_{(2)}\,.
\ee
The quadratic expressions for the current components are lengthy and we shall only discuss their schematic structure and will only focus on terms with at least two spatial derivatives. The time component of the current only contains contributions like $\M^2J_0\supset m_3^3/a^2H(\partial\pi)^2$ while 
\be
\M^2J^i=\frac{m_3^3}{a^4}\left[\partial_i\pi\Delta\pi-\partial_i\pi\partial_i\partial_j\pi\right]\,.
\ee
In particular, we see that in $J^0$ there are no terms with more than two spatial derivatives. In the quasi-static subhorizon limit, it is thus clearly subdominant in comparison to the contributions from $J^i$ involving three spatial derivatives. Hence the leading contribution arises from the spatial divergence of the current, so that $\partial_iJ^i$ is the only contribution to the quadratic equation \eqref{quadr_eom} in the subhorizon limit.

\subsection{Cubic order}\label{cubic}
When looking at cubic order, we see that to leading order $\M^2J^0\sim m_3^3/a^4(\partial\pi)^2\Delta\pi$ while $\M^2J^i\sim m_3^3/a^4(\partial\pi)^2\partial_i\pi$. This means that $J^0$ and $J^i$, similarly to what happened at linear order, would again give comparable contributions in the equation of motion. Also, there is a non-vanishing cubic contribution to $\partial\mathcal L/\partial\phi\sim (m_3^3)^{\hbox{$ \cdot$}}/a^4(\partial\pi)^2\partial^2\pi$. They would all however be subdominant with respect to the quadratic terms. The reason is that for the theory \eqref{act_cov} the screening mechanism operates at cubic order in the action, \emph{i.e.} at the quadratic order in equations of motion. 

As a side remark let us briefly consider also the case with the quartic Galilleon term present in the action. In this case the scalar field equation of motion in the subhorizon limit would contain a cubic contribution of the form \cite{Kimura:2011dc}:
\be
\begin{split}
&\partial_i\left[\partial_i\pi\left((\Delta\pi)^2-(\partial_j\partial_k\pi)^2\right)\right.\\
&\left.-2\partial_j\pi\Delta\pi\partial_i\partial_j\pi+2\partial_j\pi\partial_k\partial_j\pi\partial_i\partial_k\pi\right]=\\
&=\partial_i\partial_j\left[\partial_i\pi\partial_j\pi\Delta\pi-\partial_i\pi\partial_k\pi\partial_k\partial_j\pi\right.\\
&\left.-\delta_{ij}(\partial_k\pi)^2\Delta\pi+(\partial_k\pi)^2\partial_i\partial_j\pi\right]\,.
\end{split}
\ee
With the last equality we simply wanted to point out that the equation takes the form of a double spatial divergence. In the main body of this work the same was observed to be true also at the quadratic order.

\section{Covariant stress-energy tensor}\label{app:emtcov}
For future reference let us give the fully nonlinear expression for the variation of the covariant action~\eqref{act_cov} with respect to the metric:
\begin{widetext}
\begin{align*}
\begin{split}
\frac{1}{\sqrt{-g}}&\frac{\delta S}{\delta g^{\mu\nu}}=\frac{\M^2}{2}\left(R_{\mu\nu}-\frac{1}{2}g_{\mu\nu}R+\frac{1}{2}g_{\mu\nu}\Lambda\right)
-\frac{1}{2}g_{\mu\nu}\L_{\phi}-\frac{c}{\M^4}\partial_\mu\phi\,\partial_\nu\phi\,\\
+&\frac{m_2^4}{\M^4}\partial_\mu\phi\,\partial_\nu\phi+\frac{m_2^4}{\M^8}(\partial\phi)^2\partial_\mu\phi\,\partial_\nu\phi
+\frac{3}{8}\frac{m_3^3}{\M^2}g_{\mu\nu}\Box\phi+\frac{3}{2}\frac{m_3^3}{\M^4}H\partial_\mu\phi\,\partial_\nu\phi\,\\
+&\frac{m_3^3}{2\M^6}\left(g_{\mu\nu}\Box\phi(\partial\phi)^2+g_{\mu\nu}\partial^\alpha\phi\,\partial_\alpha(\partial\phi)^2\right.
-\left.\frac{3}{2}\partial_{(\mu}\phi\,\partial_{\nu)}(\partial\phi)^2+\frac{3}{2}\Box\phi\partial_\mu\phi\,\partial_\nu\phi\right)\,\\
+&\frac{m_3^3}{4\M^{10}}\left(\partial_\mu\phi\,\partial_\nu\phi(\partial\phi)^2\Box\phi-(\partial\phi)^2\partial_{(\mu}\phi\,\partial_{\nu)}(\partial\phi)^2\right)
+\frac{m_3^3}{4\M^{10}}\left(g_{\mu\nu}(\partial\phi)^2\partial^\alpha\phi\,\partial_\alpha(\partial\phi)^2+\frac{1}{2}g_{\mu\nu}\Box\phi\left((\partial\phi)^2\right)^2\right)\\
+&\frac{1}{2}\partial_\phi m^3_3 \left(1+\frac{(\partial\phi)^2}{\M^4}\right)\left[\frac{3}{2\M^2}\left(\frac{1}{2}g_{\mu\nu}(\partial\phi)^2-\partial_\mu\phi\,\partial_\nu\phi\right)\right.
+\left.\frac{1}{2\M^6}(\partial\phi)^2\left(\frac{1}{2}g_{\mu\nu}(\partial\phi)^2-2\partial_\mu\phi\,\partial_\nu\phi\right)\right]\,.
\end{split}
\end{align*}
\end{widetext}
where as in Appendix~\ref{app:current} all the parameters $\Lambda, c, H, m_2, m_3$ all depend on the scalar field as $\Lambda=\Lambda\left(\frac{\phi}{\M^2}\right)$ etc. 
\bigskip
\bibliographystyle{apsrev4-1}

\bibliography{lib}

\begin{thebibliography}{62}%
\makeatletter
\providecommand \@ifxundefined [1]{%
 \@ifx{#1\undefined}
}%
\providecommand \@ifnum [1]{%
 \ifnum #1\expandafter \@firstoftwo
 \else \expandafter \@secondoftwo
 \fi
}%
\providecommand \@ifx [1]{%
 \ifx #1\expandafter \@firstoftwo
 \else \expandafter \@secondoftwo
 \fi
}%
\providecommand \natexlab [1]{#1}%
\providecommand \enquote  [1]{``#1''}%
\providecommand \bibnamefont  [1]{#1}%
\providecommand \bibfnamefont [1]{#1}%
\providecommand \citenamefont [1]{#1}%
\providecommand \href@noop [0]{\@secondoftwo}%
\providecommand \href [0]{\begingroup \@sanitize@url \@href}%
\providecommand \@href[1]{\@@startlink{#1}\@@href}%
\providecommand \@@href[1]{\endgroup#1\@@endlink}%
\providecommand \@sanitize@url [0]{\catcode `\\12\catcode `\$12\catcode
  `\&12\catcode `\#12\catcode `\^12\catcode `\_12\catcode `\%12\relax}%
\providecommand \@@startlink[1]{}%
\providecommand \@@endlink[0]{}%
\providecommand \url  [0]{\begingroup\@sanitize@url \@url }%
\providecommand \@url [1]{\endgroup\@href {#1}{\urlprefix }}%
\providecommand \urlprefix  [0]{URL }%
\providecommand \Eprint [0]{\href }%
\providecommand \doibase [0]{http://dx.doi.org/}%
\providecommand \selectlanguage [0]{\@gobble}%
\providecommand \bibinfo  [0]{\@secondoftwo}%
\providecommand \bibfield  [0]{\@secondoftwo}%
\providecommand \translation [1]{[#1]}%
\providecommand \BibitemOpen [0]{}%
\providecommand \bibitemStop [0]{}%
\providecommand \bibitemNoStop [0]{.\EOS\space}%
\providecommand \EOS [0]{\spacefactor3000\relax}%
\providecommand \BibitemShut  [1]{\csname bibitem#1\endcsname}%
\let\auto@bib@innerbib\@empty
\bibitem [{\citenamefont {Gubitosi}\ \emph {et~al.}(2013)\citenamefont
  {Gubitosi}, \citenamefont {Piazza},\ and\ \citenamefont
  {Vernizzi}}]{Gubitosi:2012hu}%
  \BibitemOpen
  \bibfield  {author} {\bibinfo {author} {\bibfnamefont {G.}~\bibnamefont
  {Gubitosi}}, \bibinfo {author} {\bibfnamefont {F.}~\bibnamefont {Piazza}}, \
  and\ \bibinfo {author} {\bibfnamefont {F.}~\bibnamefont {Vernizzi}},\ }\href
  {\doibase 10.1088/1475-7516/2013/02/032} {\bibfield  {journal} {\bibinfo
  {journal} {JCAP}\ }\textbf {\bibinfo {volume} {1302}},\ \bibinfo {pages}
  {032} (\bibinfo {year} {2013})},\ \bibinfo {note} {[JCAP1302,032(2013)]},\
  \Eprint {http://arxiv.org/abs/1210.0201} {arXiv:1210.0201 [hep-th]}
  \BibitemShut {NoStop}%
\bibitem [{\citenamefont {Cheung}\ \emph {et~al.}(2008)\citenamefont {Cheung},
  \citenamefont {Creminelli}, \citenamefont {Fitzpatrick}, \citenamefont
  {Kaplan},\ and\ \citenamefont {Senatore}}]{Cheung:2007st}%
  \BibitemOpen
  \bibfield  {author} {\bibinfo {author} {\bibfnamefont {C.}~\bibnamefont
  {Cheung}}, \bibinfo {author} {\bibfnamefont {P.}~\bibnamefont {Creminelli}},
  \bibinfo {author} {\bibfnamefont {A.~L.}\ \bibnamefont {Fitzpatrick}},
  \bibinfo {author} {\bibfnamefont {J.}~\bibnamefont {Kaplan}}, \ and\ \bibinfo
  {author} {\bibfnamefont {L.}~\bibnamefont {Senatore}},\ }\href {\doibase
  10.1088/1126-6708/2008/03/014} {\bibfield  {journal} {\bibinfo  {journal}
  {JHEP}\ }\textbf {\bibinfo {volume} {03}},\ \bibinfo {pages} {014} (\bibinfo
  {year} {2008})},\ \Eprint {http://arxiv.org/abs/0709.0293} {arXiv:0709.0293
  [hep-th]} \BibitemShut {NoStop}%
\bibitem [{\citenamefont {Vainshtein}(1972)}]{Vainshtein:1972sx}%
  \BibitemOpen
  \bibfield  {author} {\bibinfo {author} {\bibfnamefont {A.~I.}\ \bibnamefont
  {Vainshtein}},\ }\href {\doibase 10.1016/0370-2693(72)90147-5} {\bibfield
  {journal} {\bibinfo  {journal} {Phys. Lett.}\ }\textbf {\bibinfo {volume}
  {39B}},\ \bibinfo {pages} {393} (\bibinfo {year} {1972})}\BibitemShut
  {NoStop}%
\bibitem [{\citenamefont {Babichev}\ and\ \citenamefont
  {Deffayet}(2013)}]{Babichev:2013usa}%
  \BibitemOpen
  \bibfield  {author} {\bibinfo {author} {\bibfnamefont {E.}~\bibnamefont
  {Babichev}}\ and\ \bibinfo {author} {\bibfnamefont {C.}~\bibnamefont
  {Deffayet}},\ }\href {\doibase 10.1088/0264-9381/30/18/184001} {\bibfield
  {journal} {\bibinfo  {journal} {Class. Quant. Grav.}\ }\textbf {\bibinfo
  {volume} {30}},\ \bibinfo {pages} {184001} (\bibinfo {year} {2013})},\
  \Eprint {http://arxiv.org/abs/1304.7240} {arXiv:1304.7240 [gr-qc]}
  \BibitemShut {NoStop}%
\bibitem [{\citenamefont {Horndeski}(1974)}]{Horndeski:1974wa}%
  \BibitemOpen
  \bibfield  {author} {\bibinfo {author} {\bibfnamefont {G.~W.}\ \bibnamefont
  {Horndeski}},\ }\href {\doibase 10.1007/BF01807638} {\bibfield  {journal}
  {\bibinfo  {journal} {Int. J. Theor. Phys.}\ }\textbf {\bibinfo {volume}
  {10}},\ \bibinfo {pages} {363} (\bibinfo {year} {1974})}\BibitemShut
  {NoStop}%
\bibitem [{\citenamefont {Deffayet}\ \emph {et~al.}(2009)\citenamefont
  {Deffayet}, \citenamefont {Esposito-Farese},\ and\ \citenamefont
  {Vikman}}]{Deffayet:2009wt}%
  \BibitemOpen
  \bibfield  {author} {\bibinfo {author} {\bibfnamefont {C.}~\bibnamefont
  {Deffayet}}, \bibinfo {author} {\bibfnamefont {G.}~\bibnamefont
  {Esposito-Farese}}, \ and\ \bibinfo {author} {\bibfnamefont {A.}~\bibnamefont
  {Vikman}},\ }\href {\doibase 10.1103/PhysRevD.79.084003} {\bibfield
  {journal} {\bibinfo  {journal} {Phys. Rev.}\ }\textbf {\bibinfo {volume}
  {D79}},\ \bibinfo {pages} {084003} (\bibinfo {year} {2009})},\ \Eprint
  {http://arxiv.org/abs/0901.1314} {arXiv:0901.1314 [hep-th]} \BibitemShut
  {NoStop}%
\bibitem [{\citenamefont {Deffayet}\ \emph {et~al.}(2011)\citenamefont
  {Deffayet}, \citenamefont {Gao}, \citenamefont {Steer},\ and\ \citenamefont
  {Zahariade}}]{Deffayet:2011gz}%
  \BibitemOpen
  \bibfield  {author} {\bibinfo {author} {\bibfnamefont {C.}~\bibnamefont
  {Deffayet}}, \bibinfo {author} {\bibfnamefont {X.}~\bibnamefont {Gao}},
  \bibinfo {author} {\bibfnamefont {D.~A.}\ \bibnamefont {Steer}}, \ and\
  \bibinfo {author} {\bibfnamefont {G.}~\bibnamefont {Zahariade}},\ }\href
  {\doibase 10.1103/PhysRevD.84.064039} {\bibfield  {journal} {\bibinfo
  {journal} {Phys. Rev.}\ }\textbf {\bibinfo {volume} {D84}},\ \bibinfo {pages}
  {064039} (\bibinfo {year} {2011})},\ \Eprint {http://arxiv.org/abs/1103.3260}
  {arXiv:1103.3260 [hep-th]} \BibitemShut {NoStop}%
\bibitem [{\citenamefont {Abbott}\ \emph {et~al.}(2016)\citenamefont {Abbott}
  \emph {et~al.}}]{Abbott:2016blz}%
  \BibitemOpen
  \bibfield  {author} {\bibinfo {author} {\bibfnamefont {B.~P.}\ \bibnamefont
  {Abbott}} \emph {et~al.} (\bibinfo {collaboration} {Virgo, LIGO
  Scientific}),\ }\href {\doibase 10.1103/PhysRevLett.116.061102} {\bibfield
  {journal} {\bibinfo  {journal} {Phys. Rev. Lett.}\ }\textbf {\bibinfo
  {volume} {116}},\ \bibinfo {pages} {061102} (\bibinfo {year} {2016})},\
  \Eprint {http://arxiv.org/abs/1602.03837} {arXiv:1602.03837 [gr-qc]}
  \BibitemShut {NoStop}%
\bibitem [{\citenamefont {Abbott}\ \emph
  {et~al.}(2017{\natexlab{a}})\citenamefont {Abbott} \emph
  {et~al.}}]{TheLIGOScientific:2017qsa}%
  \BibitemOpen
  \bibfield  {author} {\bibinfo {author} {\bibfnamefont {B.}~\bibnamefont
  {Abbott}} \emph {et~al.} (\bibinfo {collaboration} {LIGO Scientific,
  Virgo}),\ }\href {\doibase 10.1103/PhysRevLett.119.161101} {\bibfield
  {journal} {\bibinfo  {journal} {Phys. Rev. Lett.}\ }\textbf {\bibinfo
  {volume} {119}},\ \bibinfo {pages} {161101} (\bibinfo {year}
  {2017}{\natexlab{a}})},\ \Eprint {http://arxiv.org/abs/1710.05832}
  {arXiv:1710.05832 [gr-qc]} \BibitemShut {NoStop}%
\bibitem [{\citenamefont {Abbott}\ \emph
  {et~al.}(2017{\natexlab{b}})\citenamefont {Abbott} \emph
  {et~al.}}]{Monitor:2017mdv}%
  \BibitemOpen
  \bibfield  {author} {\bibinfo {author} {\bibfnamefont {B.~P.}\ \bibnamefont
  {Abbott}} \emph {et~al.} (\bibinfo {collaboration} {LIGO Scientific, Virgo,
  Fermi-GBM, INTEGRAL}),\ }\href {\doibase 10.3847/2041-8213/aa920c} {\bibfield
   {journal} {\bibinfo  {journal} {Astrophys. J.}\ }\textbf {\bibinfo {volume}
  {848}},\ \bibinfo {pages} {L13} (\bibinfo {year} {2017}{\natexlab{b}})},\
  \Eprint {http://arxiv.org/abs/1710.05834} {arXiv:1710.05834 [astro-ph.HE]}
  \BibitemShut {NoStop}%
\bibitem [{\citenamefont {Creminelli}\ and\ \citenamefont
  {Vernizzi}(2017)}]{Creminelli:2017sry}%
  \BibitemOpen
  \bibfield  {author} {\bibinfo {author} {\bibfnamefont {P.}~\bibnamefont
  {Creminelli}}\ and\ \bibinfo {author} {\bibfnamefont {F.}~\bibnamefont
  {Vernizzi}},\ }\href {\doibase 10.1103/PhysRevLett.119.251302} {\bibfield
  {journal} {\bibinfo  {journal} {Phys. Rev. Lett.}\ }\textbf {\bibinfo
  {volume} {119}},\ \bibinfo {pages} {251302} (\bibinfo {year} {2017})},\
  \Eprint {http://arxiv.org/abs/1710.05877} {arXiv:1710.05877 [astro-ph.CO]}
  \BibitemShut {NoStop}%
\bibitem [{\citenamefont {Sakstein}\ and\ \citenamefont
  {Jain}(2017)}]{Sakstein:2017xjx}%
  \BibitemOpen
  \bibfield  {author} {\bibinfo {author} {\bibfnamefont {J.}~\bibnamefont
  {Sakstein}}\ and\ \bibinfo {author} {\bibfnamefont {B.}~\bibnamefont
  {Jain}},\ }\href {\doibase 10.1103/PhysRevLett.119.251303} {\bibfield
  {journal} {\bibinfo  {journal} {Phys. Rev. Lett.}\ }\textbf {\bibinfo
  {volume} {119}},\ \bibinfo {pages} {251303} (\bibinfo {year} {2017})},\
  \Eprint {http://arxiv.org/abs/1710.05893} {arXiv:1710.05893 [astro-ph.CO]}
  \BibitemShut {NoStop}%
\bibitem [{\citenamefont {Baker}\ \emph {et~al.}(2017)\citenamefont {Baker},
  \citenamefont {Bellini}, \citenamefont {Ferreira}, \citenamefont {Lagos},
  \citenamefont {Noller},\ and\ \citenamefont {Sawicki}}]{Baker:2017hug}%
  \BibitemOpen
  \bibfield  {author} {\bibinfo {author} {\bibfnamefont {T.}~\bibnamefont
  {Baker}}, \bibinfo {author} {\bibfnamefont {E.}~\bibnamefont {Bellini}},
  \bibinfo {author} {\bibfnamefont {P.~G.}\ \bibnamefont {Ferreira}}, \bibinfo
  {author} {\bibfnamefont {M.}~\bibnamefont {Lagos}}, \bibinfo {author}
  {\bibfnamefont {J.}~\bibnamefont {Noller}}, \ and\ \bibinfo {author}
  {\bibfnamefont {I.}~\bibnamefont {Sawicki}},\ }\href {\doibase
  10.1103/PhysRevLett.119.251301} {\bibfield  {journal} {\bibinfo  {journal}
  {Phys. Rev. Lett.}\ }\textbf {\bibinfo {volume} {119}},\ \bibinfo {pages}
  {251301} (\bibinfo {year} {2017})},\ \Eprint
  {http://arxiv.org/abs/1710.06394} {arXiv:1710.06394 [astro-ph.CO]}
  \BibitemShut {NoStop}%
\bibitem [{\citenamefont {Ezquiaga}\ and\ \citenamefont
  {Zumalac\'arregui}(2017)}]{Ezquiaga:2017ekz}%
  \BibitemOpen
  \bibfield  {author} {\bibinfo {author} {\bibfnamefont {J.~M.}\ \bibnamefont
  {Ezquiaga}}\ and\ \bibinfo {author} {\bibfnamefont {M.}~\bibnamefont
  {Zumalac\'arregui}},\ }\href {\doibase 10.1103/PhysRevLett.119.251304}
  {\bibfield  {journal} {\bibinfo  {journal} {Phys. Rev. Lett.}\ }\textbf
  {\bibinfo {volume} {119}},\ \bibinfo {pages} {251304} (\bibinfo {year}
  {2017})},\ \Eprint {http://arxiv.org/abs/1710.05901} {arXiv:1710.05901
  [astro-ph.CO]} \BibitemShut {NoStop}%
\bibitem [{\citenamefont {de~Rham}\ and\ \citenamefont
  {Melville}(2018)}]{deRham:2018red}%
  \BibitemOpen
  \bibfield  {author} {\bibinfo {author} {\bibfnamefont {C.}~\bibnamefont
  {de~Rham}}\ and\ \bibinfo {author} {\bibfnamefont {S.}~\bibnamefont
  {Melville}},\ }\href@noop {} {\  (\bibinfo {year} {2018})},\ \Eprint
  {http://arxiv.org/abs/1806.09417} {arXiv:1806.09417 [hep-th]} \BibitemShut
  {NoStop}%
\bibitem [{\citenamefont {Dong}\ \emph {et~al.}(2017)\citenamefont {Dong},
  \citenamefont {Sakstein},\ and\ \citenamefont {Stojkovic}}]{Dong:2017toi}%
  \BibitemOpen
  \bibfield  {author} {\bibinfo {author} {\bibfnamefont {R.}~\bibnamefont
  {Dong}}, \bibinfo {author} {\bibfnamefont {J.}~\bibnamefont {Sakstein}}, \
  and\ \bibinfo {author} {\bibfnamefont {D.}~\bibnamefont {Stojkovic}},\ }\href
  {\doibase 10.1103/PhysRevD.96.064048} {\bibfield  {journal} {\bibinfo
  {journal} {Phys. Rev.}\ }\textbf {\bibinfo {volume} {D96}},\ \bibinfo {pages}
  {064048} (\bibinfo {year} {2017})},\ \Eprint
  {http://arxiv.org/abs/1709.01641} {arXiv:1709.01641 [gr-qc]} \BibitemShut
  {NoStop}%
\bibitem [{\citenamefont {Tattersall}\ and\ \citenamefont
  {Ferreira}(2018)}]{Tattersall:2018nve}%
  \BibitemOpen
  \bibfield  {author} {\bibinfo {author} {\bibfnamefont {O.~J.}\ \bibnamefont
  {Tattersall}}\ and\ \bibinfo {author} {\bibfnamefont {P.~G.}\ \bibnamefont
  {Ferreira}},\ }\href {\doibase 10.1103/PhysRevD.97.104047} {\bibfield
  {journal} {\bibinfo  {journal} {Phys. Rev.}\ }\textbf {\bibinfo {volume}
  {D97}},\ \bibinfo {pages} {104047} (\bibinfo {year} {2018})},\ \Eprint
  {http://arxiv.org/abs/1804.08950} {arXiv:1804.08950 [gr-qc]} \BibitemShut
  {NoStop}%
\bibitem [{\citenamefont {Franciolini}\ \emph {et~al.}(2018)\citenamefont
  {Franciolini}, \citenamefont {Hui}, \citenamefont {Penco}, \citenamefont
  {Santoni},\ and\ \citenamefont {Trincherini}}]{Franciolini:2018uyq}%
  \BibitemOpen
  \bibfield  {author} {\bibinfo {author} {\bibfnamefont {G.}~\bibnamefont
  {Franciolini}}, \bibinfo {author} {\bibfnamefont {L.}~\bibnamefont {Hui}},
  \bibinfo {author} {\bibfnamefont {R.}~\bibnamefont {Penco}}, \bibinfo
  {author} {\bibfnamefont {L.}~\bibnamefont {Santoni}}, \ and\ \bibinfo
  {author} {\bibfnamefont {E.}~\bibnamefont {Trincherini}},\ }\href@noop {} {\
  (\bibinfo {year} {2018})},\ \Eprint {http://arxiv.org/abs/1810.07706}
  {arXiv:1810.07706 [hep-th]} \BibitemShut {NoStop}%
\bibitem [{\citenamefont {Witek}\ \emph {et~al.}(2018)\citenamefont {Witek},
  \citenamefont {Gualtieri}, \citenamefont {Pani},\ and\ \citenamefont
  {Sotiriou}}]{Witek:2018dmd}%
  \BibitemOpen
  \bibfield  {author} {\bibinfo {author} {\bibfnamefont {H.}~\bibnamefont
  {Witek}}, \bibinfo {author} {\bibfnamefont {L.}~\bibnamefont {Gualtieri}},
  \bibinfo {author} {\bibfnamefont {P.}~\bibnamefont {Pani}}, \ and\ \bibinfo
  {author} {\bibfnamefont {T.~P.}\ \bibnamefont {Sotiriou}},\ }\href@noop {} {\
   (\bibinfo {year} {2018})},\ \Eprint {http://arxiv.org/abs/1810.05177}
  {arXiv:1810.05177 [gr-qc]} \BibitemShut {NoStop}%
\bibitem [{\citenamefont {Mirbabayi}(2018)}]{Mirbabayi:2018mdm}%
  \BibitemOpen
  \bibfield  {author} {\bibinfo {author} {\bibfnamefont {M.}~\bibnamefont
  {Mirbabayi}},\ }\href@noop {} {\  (\bibinfo {year} {2018})},\ \Eprint
  {http://arxiv.org/abs/1807.04843} {arXiv:1807.04843 [gr-qc]} \BibitemShut
  {NoStop}%
\bibitem [{\citenamefont {Berti}\ \emph {et~al.}(2015)\citenamefont {Berti}
  \emph {et~al.}}]{Berti:2015itd}%
  \BibitemOpen
  \bibfield  {author} {\bibinfo {author} {\bibfnamefont {E.}~\bibnamefont
  {Berti}} \emph {et~al.},\ }\href {\doibase 10.1088/0264-9381/32/24/243001}
  {\bibfield  {journal} {\bibinfo  {journal} {Class. Quant. Grav.}\ }\textbf
  {\bibinfo {volume} {32}},\ \bibinfo {pages} {243001} (\bibinfo {year}
  {2015})},\ \Eprint {http://arxiv.org/abs/1501.07274} {arXiv:1501.07274
  [gr-qc]} \BibitemShut {NoStop}%
\bibitem [{\citenamefont {Will}(2006)}]{Will:2005va}%
  \BibitemOpen
  \bibfield  {author} {\bibinfo {author} {\bibfnamefont {C.~M.}\ \bibnamefont
  {Will}},\ }\href {\doibase 10.12942/lrr-2006-3} {\bibfield  {journal}
  {\bibinfo  {journal} {Living Rev. Rel.}\ }\textbf {\bibinfo {volume} {9}},\
  \bibinfo {pages} {3} (\bibinfo {year} {2006})},\ \Eprint
  {http://arxiv.org/abs/gr-qc/0510072} {arXiv:gr-qc/0510072 [gr-qc]}
  \BibitemShut {NoStop}%
\bibitem [{\citenamefont {Hui}\ \emph {et~al.}(2009)\citenamefont {Hui},
  \citenamefont {Nicolis},\ and\ \citenamefont {Stubbs}}]{Hui:2009kc}%
  \BibitemOpen
  \bibfield  {author} {\bibinfo {author} {\bibfnamefont {L.}~\bibnamefont
  {Hui}}, \bibinfo {author} {\bibfnamefont {A.}~\bibnamefont {Nicolis}}, \ and\
  \bibinfo {author} {\bibfnamefont {C.}~\bibnamefont {Stubbs}},\ }\href
  {\doibase 10.1103/PhysRevD.80.104002} {\bibfield  {journal} {\bibinfo
  {journal} {Phys. Rev.}\ }\textbf {\bibinfo {volume} {D80}},\ \bibinfo {pages}
  {104002} (\bibinfo {year} {2009})},\ \Eprint {http://arxiv.org/abs/0905.2966}
  {arXiv:0905.2966 [astro-ph.CO]} \BibitemShut {NoStop}%
\bibitem [{\citenamefont {Nordtvedt}(1968)}]{Nordtvedt:1968qr}%
  \BibitemOpen
  \bibfield  {author} {\bibinfo {author} {\bibfnamefont {K.}~\bibnamefont
  {Nordtvedt}},\ }\href {\doibase 10.1103/PhysRev.169.1014} {\bibfield
  {journal} {\bibinfo  {journal} {Phys. Rev.}\ }\textbf {\bibinfo {volume}
  {169}},\ \bibinfo {pages} {1014} (\bibinfo {year} {1968})}\BibitemShut
  {NoStop}%
\bibitem [{\citenamefont {Barausse}(2017)}]{Barausse:2017gip}%
  \BibitemOpen
  \bibfield  {author} {\bibinfo {author} {\bibfnamefont {E.}~\bibnamefont
  {Barausse}},\ }\bibfield  {booktitle} {\emph {\bibinfo {booktitle}
  {{Proceedings, 3rd International Symposium on Quest for the Origin of
  Particles and the Universe (KMI2017): Nagoya, Japan, January 5-7, 2017}}},\
  }\href {\doibase 10.22323/1.294.0029} {\bibfield  {journal} {\bibinfo
  {journal} {PoS}\ }\textbf {\bibinfo {volume} {KMI2017}},\ \bibinfo {pages}
  {029} (\bibinfo {year} {2017})},\ \Eprint {http://arxiv.org/abs/1703.05699}
  {arXiv:1703.05699 [gr-qc]} \BibitemShut {NoStop}%
\bibitem [{\citenamefont {Freire}\ \emph {et~al.}(2012)\citenamefont {Freire},
  \citenamefont {Wex}, \citenamefont {Esposito-Farese}, \citenamefont
  {Verbiest}, \citenamefont {Bailes}, \citenamefont {Jacoby}, \citenamefont
  {Kramer}, \citenamefont {Stairs}, \citenamefont {Antoniadis},\ and\
  \citenamefont {Janssen}}]{Freire:2012mg}%
  \BibitemOpen
  \bibfield  {author} {\bibinfo {author} {\bibfnamefont {P.~C.~C.}\
  \bibnamefont {Freire}}, \bibinfo {author} {\bibfnamefont {N.}~\bibnamefont
  {Wex}}, \bibinfo {author} {\bibfnamefont {G.}~\bibnamefont
  {Esposito-Farese}}, \bibinfo {author} {\bibfnamefont {J.~P.~W.}\ \bibnamefont
  {Verbiest}}, \bibinfo {author} {\bibfnamefont {M.}~\bibnamefont {Bailes}},
  \bibinfo {author} {\bibfnamefont {B.~A.}\ \bibnamefont {Jacoby}}, \bibinfo
  {author} {\bibfnamefont {M.}~\bibnamefont {Kramer}}, \bibinfo {author}
  {\bibfnamefont {I.~H.}\ \bibnamefont {Stairs}}, \bibinfo {author}
  {\bibfnamefont {J.}~\bibnamefont {Antoniadis}}, \ and\ \bibinfo {author}
  {\bibfnamefont {G.~H.}\ \bibnamefont {Janssen}},\ }\href {\doibase
  10.1111/j.1365-2966.2012.21253.x} {\bibfield  {journal} {\bibinfo  {journal}
  {Mon. Not. Roy. Astron. Soc.}\ }\textbf {\bibinfo {volume} {423}},\ \bibinfo
  {pages} {3328} (\bibinfo {year} {2012})},\ \Eprint
  {http://arxiv.org/abs/1205.1450} {arXiv:1205.1450 [astro-ph.GA]} \BibitemShut
  {NoStop}%
\bibitem [{\citenamefont {Brans}\ and\ \citenamefont
  {Dicke}(1961)}]{Brans:1961sx}%
  \BibitemOpen
  \bibfield  {author} {\bibinfo {author} {\bibfnamefont {C.}~\bibnamefont
  {Brans}}\ and\ \bibinfo {author} {\bibfnamefont {R.~H.}\ \bibnamefont
  {Dicke}},\ }\href {\doibase 10.1103/PhysRev.124.925} {\bibfield  {journal}
  {\bibinfo  {journal} {Phys. Rev.}\ }\textbf {\bibinfo {volume} {124}},\
  \bibinfo {pages} {925} (\bibinfo {year} {1961})}\BibitemShut {NoStop}%
\bibitem [{\citenamefont {Bekenstein}(1972)}]{Bekenstein:1971hc}%
  \BibitemOpen
  \bibfield  {author} {\bibinfo {author} {\bibfnamefont {J.~D.}\ \bibnamefont
  {Bekenstein}},\ }\href {\doibase 10.1103/PhysRevD.5.1239} {\bibfield
  {journal} {\bibinfo  {journal} {Phys. Rev.}\ }\textbf {\bibinfo {volume}
  {D5}},\ \bibinfo {pages} {1239} (\bibinfo {year} {1972})}\BibitemShut
  {NoStop}%
\bibitem [{\citenamefont {Bekenstein}(1995)}]{Bekenstein:1995un}%
  \BibitemOpen
  \bibfield  {author} {\bibinfo {author} {\bibfnamefont {J.~D.}\ \bibnamefont
  {Bekenstein}},\ }\href {\doibase 10.1103/PhysRevD.51.R6608} {\bibfield
  {journal} {\bibinfo  {journal} {Phys. Rev.}\ }\textbf {\bibinfo {volume}
  {D51}},\ \bibinfo {pages} {R6608} (\bibinfo {year} {1995})}\BibitemShut
  {NoStop}%
\bibitem [{\citenamefont {Hui}\ and\ \citenamefont
  {Nicolis}(2013)}]{Hui:2012qt}%
  \BibitemOpen
  \bibfield  {author} {\bibinfo {author} {\bibfnamefont {L.}~\bibnamefont
  {Hui}}\ and\ \bibinfo {author} {\bibfnamefont {A.}~\bibnamefont {Nicolis}},\
  }\href {\doibase 10.1103/PhysRevLett.110.241104} {\bibfield  {journal}
  {\bibinfo  {journal} {Phys. Rev. Lett.}\ }\textbf {\bibinfo {volume} {110}},\
  \bibinfo {pages} {241104} (\bibinfo {year} {2013})},\ \Eprint
  {http://arxiv.org/abs/1202.1296} {arXiv:1202.1296 [hep-th]} \BibitemShut
  {NoStop}%
\bibitem [{\citenamefont {Nicolis}\ \emph {et~al.}(2009)\citenamefont
  {Nicolis}, \citenamefont {Rattazzi},\ and\ \citenamefont
  {Trincherini}}]{Nicolis:2008in}%
  \BibitemOpen
  \bibfield  {author} {\bibinfo {author} {\bibfnamefont {A.}~\bibnamefont
  {Nicolis}}, \bibinfo {author} {\bibfnamefont {R.}~\bibnamefont {Rattazzi}}, \
  and\ \bibinfo {author} {\bibfnamefont {E.}~\bibnamefont {Trincherini}},\
  }\href {\doibase 10.1103/PhysRevD.79.064036} {\bibfield  {journal} {\bibinfo
  {journal} {Phys. Rev.}\ }\textbf {\bibinfo {volume} {D79}},\ \bibinfo {pages}
  {064036} (\bibinfo {year} {2009})},\ \Eprint {http://arxiv.org/abs/0811.2197}
  {arXiv:0811.2197 [hep-th]} \BibitemShut {NoStop}%
\bibitem [{\citenamefont {Sotiriou}\ and\ \citenamefont
  {Zhou}(2014{\natexlab{a}})}]{Sotiriou:2013qea}%
  \BibitemOpen
  \bibfield  {author} {\bibinfo {author} {\bibfnamefont {T.~P.}\ \bibnamefont
  {Sotiriou}}\ and\ \bibinfo {author} {\bibfnamefont {S.-Y.}\ \bibnamefont
  {Zhou}},\ }\href {\doibase 10.1103/PhysRevLett.112.251102} {\bibfield
  {journal} {\bibinfo  {journal} {Phys. Rev. Lett.}\ }\textbf {\bibinfo
  {volume} {112}},\ \bibinfo {pages} {251102} (\bibinfo {year}
  {2014}{\natexlab{a}})},\ \Eprint {http://arxiv.org/abs/1312.3622}
  {arXiv:1312.3622 [gr-qc]} \BibitemShut {NoStop}%
\bibitem [{\citenamefont {Sotiriou}\ and\ \citenamefont
  {Zhou}(2014{\natexlab{b}})}]{Sotiriou:2014pfa}%
  \BibitemOpen
  \bibfield  {author} {\bibinfo {author} {\bibfnamefont {T.~P.}\ \bibnamefont
  {Sotiriou}}\ and\ \bibinfo {author} {\bibfnamefont {S.-Y.}\ \bibnamefont
  {Zhou}},\ }\href {\doibase 10.1103/PhysRevD.90.124063} {\bibfield  {journal}
  {\bibinfo  {journal} {Phys. Rev.}\ }\textbf {\bibinfo {volume} {D90}},\
  \bibinfo {pages} {124063} (\bibinfo {year} {2014}{\natexlab{b}})},\ \Eprint
  {http://arxiv.org/abs/1408.1698} {arXiv:1408.1698 [gr-qc]} \BibitemShut
  {NoStop}%
\bibitem [{\citenamefont {Lehebel}\ \emph {et~al.}(2017)\citenamefont
  {Lehebel}, \citenamefont {Babichev},\ and\ \citenamefont
  {Charmousis}}]{Lehebel:2017fag}%
  \BibitemOpen
  \bibfield  {author} {\bibinfo {author} {\bibfnamefont {A.}~\bibnamefont
  {Lehebel}}, \bibinfo {author} {\bibfnamefont {E.}~\bibnamefont {Babichev}}, \
  and\ \bibinfo {author} {\bibfnamefont {C.}~\bibnamefont {Charmousis}},\
  }\href {\doibase 10.1088/1475-7516/2017/07/037} {\bibfield  {journal}
  {\bibinfo  {journal} {JCAP}\ }\textbf {\bibinfo {volume} {1707}},\ \bibinfo
  {pages} {037} (\bibinfo {year} {2017})},\ \Eprint
  {http://arxiv.org/abs/1706.04989} {arXiv:1706.04989 [gr-qc]} \BibitemShut
  {NoStop}%
\bibitem [{\citenamefont {Barausse}\ and\ \citenamefont
  {Yagi}(2015)}]{Barausse:2015wia}%
  \BibitemOpen
  \bibfield  {author} {\bibinfo {author} {\bibfnamefont {E.}~\bibnamefont
  {Barausse}}\ and\ \bibinfo {author} {\bibfnamefont {K.}~\bibnamefont
  {Yagi}},\ }\href {\doibase 10.1103/PhysRevLett.115.211105} {\bibfield
  {journal} {\bibinfo  {journal} {Phys. Rev. Lett.}\ }\textbf {\bibinfo
  {volume} {115}},\ \bibinfo {pages} {211105} (\bibinfo {year} {2015})},\
  \Eprint {http://arxiv.org/abs/1509.04539} {arXiv:1509.04539 [gr-qc]}
  \BibitemShut {NoStop}%
\bibitem [{\citenamefont {Rinaldi}(2012)}]{Rinaldi:2012vy}%
  \BibitemOpen
  \bibfield  {author} {\bibinfo {author} {\bibfnamefont {M.}~\bibnamefont
  {Rinaldi}},\ }\href {\doibase 10.1103/PhysRevD.86.084048} {\bibfield
  {journal} {\bibinfo  {journal} {Phys. Rev.}\ }\textbf {\bibinfo {volume}
  {D86}},\ \bibinfo {pages} {084048} (\bibinfo {year} {2012})},\ \Eprint
  {http://arxiv.org/abs/1208.0103} {arXiv:1208.0103 [gr-qc]} \BibitemShut
  {NoStop}%
\bibitem [{\citenamefont {Minamitsuji}(2014)}]{Minamitsuji:2013ura}%
  \BibitemOpen
  \bibfield  {author} {\bibinfo {author} {\bibfnamefont {M.}~\bibnamefont
  {Minamitsuji}},\ }\href {\doibase 10.1103/PhysRevD.89.064017} {\bibfield
  {journal} {\bibinfo  {journal} {Phys. Rev.}\ }\textbf {\bibinfo {volume}
  {D89}},\ \bibinfo {pages} {064017} (\bibinfo {year} {2014})},\ \Eprint
  {http://arxiv.org/abs/1312.3759} {arXiv:1312.3759 [gr-qc]} \BibitemShut
  {NoStop}%
\bibitem [{\citenamefont {Charmousis}\ \emph {et~al.}(2012)\citenamefont
  {Charmousis}, \citenamefont {Copeland}, \citenamefont {Padilla},\ and\
  \citenamefont {Saffin}}]{Charmousis:2011bf}%
  \BibitemOpen
  \bibfield  {author} {\bibinfo {author} {\bibfnamefont {C.}~\bibnamefont
  {Charmousis}}, \bibinfo {author} {\bibfnamefont {E.~J.}\ \bibnamefont
  {Copeland}}, \bibinfo {author} {\bibfnamefont {A.}~\bibnamefont {Padilla}}, \
  and\ \bibinfo {author} {\bibfnamefont {P.~M.}\ \bibnamefont {Saffin}},\
  }\href {\doibase 10.1103/PhysRevLett.108.051101} {\bibfield  {journal}
  {\bibinfo  {journal} {Phys. Rev. Lett.}\ }\textbf {\bibinfo {volume} {108}},\
  \bibinfo {pages} {051101} (\bibinfo {year} {2012})},\ \Eprint
  {http://arxiv.org/abs/1106.2000} {arXiv:1106.2000 [hep-th]} \BibitemShut
  {NoStop}%
\bibitem [{\citenamefont {Herdeiro}\ and\ \citenamefont
  {Radu}(2015)}]{Herdeiro:2015waa}%
  \BibitemOpen
  \bibfield  {author} {\bibinfo {author} {\bibfnamefont {C.~A.~R.}\
  \bibnamefont {Herdeiro}}\ and\ \bibinfo {author} {\bibfnamefont
  {E.}~\bibnamefont {Radu}},\ }\bibfield  {booktitle} {\emph {\bibinfo
  {booktitle} {{Proceedings, 7th Black Holes Workshop 2014: Aveiro, Portugal,
  December 18-19, 2014}}},\ }\href {\doibase 10.1142/S0218271815420146}
  {\bibfield  {journal} {\bibinfo  {journal} {Int. J. Mod. Phys.}\ }\textbf
  {\bibinfo {volume} {D24}},\ \bibinfo {pages} {1542014} (\bibinfo {year}
  {2015})},\ \Eprint {http://arxiv.org/abs/1504.08209} {arXiv:1504.08209
  [gr-qc]} \BibitemShut {NoStop}%
\bibitem [{\citenamefont {Babichev}\ and\ \citenamefont
  {Esposito-Far\`ese}(2013)}]{Babichev:2012re}%
  \BibitemOpen
  \bibfield  {author} {\bibinfo {author} {\bibfnamefont {E.}~\bibnamefont
  {Babichev}}\ and\ \bibinfo {author} {\bibfnamefont {G.}~\bibnamefont
  {Esposito-Far\`ese}},\ }\href {\doibase 10.1103/PhysRevD.87.044032}
  {\bibfield  {journal} {\bibinfo  {journal} {Phys. Rev.}\ }\textbf {\bibinfo
  {volume} {D87}},\ \bibinfo {pages} {044032} (\bibinfo {year} {2013})},\
  \Eprint {http://arxiv.org/abs/1212.1394} {arXiv:1212.1394 [gr-qc]}
  \BibitemShut {NoStop}%
\bibitem [{\citenamefont {Babichev}\ and\ \citenamefont
  {Charmousis}(2014)}]{Babichev:2013cya}%
  \BibitemOpen
  \bibfield  {author} {\bibinfo {author} {\bibfnamefont {E.}~\bibnamefont
  {Babichev}}\ and\ \bibinfo {author} {\bibfnamefont {C.}~\bibnamefont
  {Charmousis}},\ }\href {\doibase 10.1007/JHEP08(2014)106} {\bibfield
  {journal} {\bibinfo  {journal} {JHEP}\ }\textbf {\bibinfo {volume} {08}},\
  \bibinfo {pages} {106} (\bibinfo {year} {2014})},\ \Eprint
  {http://arxiv.org/abs/1312.3204} {arXiv:1312.3204 [gr-qc]} \BibitemShut
  {NoStop}%
\bibitem [{\citenamefont {Cisterna}\ \emph {et~al.}(2015)\citenamefont
  {Cisterna}, \citenamefont {Delsate},\ and\ \citenamefont
  {Rinaldi}}]{Cisterna:2015yla}%
  \BibitemOpen
  \bibfield  {author} {\bibinfo {author} {\bibfnamefont {A.}~\bibnamefont
  {Cisterna}}, \bibinfo {author} {\bibfnamefont {T.}~\bibnamefont {Delsate}}, \
  and\ \bibinfo {author} {\bibfnamefont {M.}~\bibnamefont {Rinaldi}},\ }\href
  {\doibase 10.1103/PhysRevD.92.044050} {\bibfield  {journal} {\bibinfo
  {journal} {Phys. Rev.}\ }\textbf {\bibinfo {volume} {D92}},\ \bibinfo {pages}
  {044050} (\bibinfo {year} {2015})},\ \Eprint
  {http://arxiv.org/abs/1504.05189} {arXiv:1504.05189 [gr-qc]} \BibitemShut
  {NoStop}%
\bibitem [{\citenamefont {Babichev}\ \emph {et~al.}(2016)\citenamefont
  {Babichev}, \citenamefont {Charmousis}, \citenamefont {Lehebel},\ and\
  \citenamefont {Moskalets}}]{Babichev:2016fbg}%
  \BibitemOpen
  \bibfield  {author} {\bibinfo {author} {\bibfnamefont {E.}~\bibnamefont
  {Babichev}}, \bibinfo {author} {\bibfnamefont {C.}~\bibnamefont
  {Charmousis}}, \bibinfo {author} {\bibfnamefont {A.}~\bibnamefont {Lehebel}},
  \ and\ \bibinfo {author} {\bibfnamefont {T.}~\bibnamefont {Moskalets}},\
  }\href {\doibase 10.1088/1475-7516/2016/09/011} {\bibfield  {journal}
  {\bibinfo  {journal} {JCAP}\ }\textbf {\bibinfo {volume} {1609}},\ \bibinfo
  {pages} {011} (\bibinfo {year} {2016})},\ \Eprint
  {http://arxiv.org/abs/1605.07438} {arXiv:1605.07438 [gr-qc]} \BibitemShut
  {NoStop}%
\bibitem [{\citenamefont {Hui}\ and\ \citenamefont
  {Nicolis}(2010)}]{Hui:2010dn}%
  \BibitemOpen
  \bibfield  {author} {\bibinfo {author} {\bibfnamefont {L.}~\bibnamefont
  {Hui}}\ and\ \bibinfo {author} {\bibfnamefont {A.}~\bibnamefont {Nicolis}},\
  }\href {\doibase 10.1103/PhysRevLett.105.231101} {\bibfield  {journal}
  {\bibinfo  {journal} {Phys. Rev. Lett.}\ }\textbf {\bibinfo {volume} {105}},\
  \bibinfo {pages} {231101} (\bibinfo {year} {2010})},\ \Eprint
  {http://arxiv.org/abs/1009.2520} {arXiv:1009.2520 [hep-th]} \BibitemShut
  {NoStop}%
\bibitem [{\citenamefont {Khoury}\ and\ \citenamefont
  {Weltman}(2004{\natexlab{a}})}]{Khoury:2003aq}%
  \BibitemOpen
  \bibfield  {author} {\bibinfo {author} {\bibfnamefont {J.}~\bibnamefont
  {Khoury}}\ and\ \bibinfo {author} {\bibfnamefont {A.}~\bibnamefont
  {Weltman}},\ }\href {\doibase 10.1103/PhysRevLett.93.171104} {\bibfield
  {journal} {\bibinfo  {journal} {Phys. Rev. Lett.}\ }\textbf {\bibinfo
  {volume} {93}},\ \bibinfo {pages} {171104} (\bibinfo {year}
  {2004}{\natexlab{a}})},\ \Eprint {http://arxiv.org/abs/astro-ph/0309300}
  {arXiv:astro-ph/0309300 [astro-ph]} \BibitemShut {NoStop}%
\bibitem [{\citenamefont {Khoury}\ and\ \citenamefont
  {Weltman}(2004{\natexlab{b}})}]{Khoury:2003rn}%
  \BibitemOpen
  \bibfield  {author} {\bibinfo {author} {\bibfnamefont {J.}~\bibnamefont
  {Khoury}}\ and\ \bibinfo {author} {\bibfnamefont {A.}~\bibnamefont
  {Weltman}},\ }\href {\doibase 10.1103/PhysRevD.69.044026} {\bibfield
  {journal} {\bibinfo  {journal} {Phys. Rev.}\ }\textbf {\bibinfo {volume}
  {D69}},\ \bibinfo {pages} {044026} (\bibinfo {year} {2004}{\natexlab{b}})},\
  \Eprint {http://arxiv.org/abs/astro-ph/0309411} {arXiv:astro-ph/0309411
  [astro-ph]} \BibitemShut {NoStop}%
\bibitem [{\citenamefont {Einstein}\ \emph {et~al.}(1938)\citenamefont
  {Einstein}, \citenamefont {Infeld},\ and\ \citenamefont
  {Hoffmann}}]{Einstein:1938yz}%
  \BibitemOpen
  \bibfield  {author} {\bibinfo {author} {\bibfnamefont {A.}~\bibnamefont
  {Einstein}}, \bibinfo {author} {\bibfnamefont {L.}~\bibnamefont {Infeld}}, \
  and\ \bibinfo {author} {\bibfnamefont {B.}~\bibnamefont {Hoffmann}},\ }\href
  {\doibase 10.2307/1968714} {\bibfield  {journal} {\bibinfo  {journal} {Annals
  Math.}\ }\textbf {\bibinfo {volume} {39}},\ \bibinfo {pages} {65} (\bibinfo
  {year} {1938})}\BibitemShut {NoStop}%
\bibitem [{\citenamefont {Damour}(1989)}]{damour}%
  \BibitemOpen
  \bibfield  {author} {\bibinfo {author} {\bibfnamefont {T.}~\bibnamefont
  {Damour}},\ }\href@noop {} {\emph {\bibinfo {title} {Three Hundred Years of
  Gravitation}}}\ (\bibinfo  {publisher} {UK: Univ. Pr.},\ \bibinfo {address}
  {Cambridge},\ \bibinfo {year} {1989})\ p.\ \bibinfo {pages} {128}\BibitemShut
  {NoStop}%
\bibitem [{\citenamefont {de~Rham}\ \emph {et~al.}(2013)\citenamefont
  {de~Rham}, \citenamefont {Tolley},\ and\ \citenamefont
  {Wesley}}]{deRham:2012fw}%
  \BibitemOpen
  \bibfield  {author} {\bibinfo {author} {\bibfnamefont {C.}~\bibnamefont
  {de~Rham}}, \bibinfo {author} {\bibfnamefont {A.~J.}\ \bibnamefont {Tolley}},
  \ and\ \bibinfo {author} {\bibfnamefont {D.~H.}\ \bibnamefont {Wesley}},\
  }\href {\doibase 10.1103/PhysRevD.87.044025} {\bibfield  {journal} {\bibinfo
  {journal} {Phys. Rev.}\ }\textbf {\bibinfo {volume} {D87}},\ \bibinfo {pages}
  {044025} (\bibinfo {year} {2013})},\ \Eprint {http://arxiv.org/abs/1208.0580}
  {arXiv:1208.0580 [gr-qc]} \BibitemShut {NoStop}%
\bibitem [{\citenamefont {Gleyzes}\ \emph {et~al.}(2015)\citenamefont
  {Gleyzes}, \citenamefont {Langlois}, \citenamefont {Piazza},\ and\
  \citenamefont {Vernizzi}}]{Gleyzes:2014dya}%
  \BibitemOpen
  \bibfield  {author} {\bibinfo {author} {\bibfnamefont {J.}~\bibnamefont
  {Gleyzes}}, \bibinfo {author} {\bibfnamefont {D.}~\bibnamefont {Langlois}},
  \bibinfo {author} {\bibfnamefont {F.}~\bibnamefont {Piazza}}, \ and\ \bibinfo
  {author} {\bibfnamefont {F.}~\bibnamefont {Vernizzi}},\ }\href {\doibase
  10.1103/PhysRevLett.114.211101} {\bibfield  {journal} {\bibinfo  {journal}
  {Phys. Rev. Lett.}\ }\textbf {\bibinfo {volume} {114}},\ \bibinfo {pages}
  {211101} (\bibinfo {year} {2015})},\ \Eprint {http://arxiv.org/abs/1404.6495}
  {arXiv:1404.6495 [hep-th]} \BibitemShut {NoStop}%
\bibitem [{\citenamefont {Creminelli}\ \emph {et~al.}(2018)\citenamefont
  {Creminelli}, \citenamefont {Lewandowski}, \citenamefont {Tambalo},\ and\
  \citenamefont {Vernizzi}}]{Creminelli:2018xsv}%
  \BibitemOpen
  \bibfield  {author} {\bibinfo {author} {\bibfnamefont {P.}~\bibnamefont
  {Creminelli}}, \bibinfo {author} {\bibfnamefont {M.}~\bibnamefont
  {Lewandowski}}, \bibinfo {author} {\bibfnamefont {G.}~\bibnamefont
  {Tambalo}}, \ and\ \bibinfo {author} {\bibfnamefont {F.}~\bibnamefont
  {Vernizzi}},\ }\href@noop {} {\  (\bibinfo {year} {2018})},\ \Eprint
  {http://arxiv.org/abs/1809.03484} {arXiv:1809.03484 [astro-ph.CO]}
  \BibitemShut {NoStop}%
\bibitem [{\citenamefont {Hawking}(1972)}]{Hawking:1972qk}%
  \BibitemOpen
  \bibfield  {author} {\bibinfo {author} {\bibfnamefont {S.~W.}\ \bibnamefont
  {Hawking}},\ }\href {\doibase 10.1007/BF01877518} {\bibfield  {journal}
  {\bibinfo  {journal} {Commun. Math. Phys.}\ }\textbf {\bibinfo {volume}
  {25}},\ \bibinfo {pages} {167} (\bibinfo {year} {1972})}\BibitemShut
  {NoStop}%
\bibitem [{\citenamefont {Arkani-Hamed}\ \emph {et~al.}(2003)\citenamefont
  {Arkani-Hamed}, \citenamefont {Georgi},\ and\ \citenamefont
  {Schwartz}}]{ArkaniHamed:2002sp}%
  \BibitemOpen
  \bibfield  {author} {\bibinfo {author} {\bibfnamefont {N.}~\bibnamefont
  {Arkani-Hamed}}, \bibinfo {author} {\bibfnamefont {H.}~\bibnamefont
  {Georgi}}, \ and\ \bibinfo {author} {\bibfnamefont {M.~D.}\ \bibnamefont
  {Schwartz}},\ }\href {\doibase 10.1016/S0003-4916(03)00068-X} {\bibfield
  {journal} {\bibinfo  {journal} {Annals Phys.}\ }\textbf {\bibinfo {volume}
  {305}},\ \bibinfo {pages} {96} (\bibinfo {year} {2003})},\ \Eprint
  {http://arxiv.org/abs/hep-th/0210184} {arXiv:hep-th/0210184 [hep-th]}
  \BibitemShut {NoStop}%
\bibitem [{\citenamefont {de~Rham}\ and\ \citenamefont
  {Gabadadze}(2010)}]{deRham:2010ik}%
  \BibitemOpen
  \bibfield  {author} {\bibinfo {author} {\bibfnamefont {C.}~\bibnamefont
  {de~Rham}}\ and\ \bibinfo {author} {\bibfnamefont {G.}~\bibnamefont
  {Gabadadze}},\ }\href {\doibase 10.1103/PhysRevD.82.044020} {\bibfield
  {journal} {\bibinfo  {journal} {Phys. Rev.}\ }\textbf {\bibinfo {volume}
  {D82}},\ \bibinfo {pages} {044020} (\bibinfo {year} {2010})},\ \Eprint
  {http://arxiv.org/abs/1007.0443} {arXiv:1007.0443 [hep-th]} \BibitemShut
  {NoStop}%
\bibitem [{\citenamefont {de~Rham}\ \emph {et~al.}(2011)\citenamefont
  {de~Rham}, \citenamefont {Gabadadze},\ and\ \citenamefont
  {Tolley}}]{deRham:2010kj}%
  \BibitemOpen
  \bibfield  {author} {\bibinfo {author} {\bibfnamefont {C.}~\bibnamefont
  {de~Rham}}, \bibinfo {author} {\bibfnamefont {G.}~\bibnamefont {Gabadadze}},
  \ and\ \bibinfo {author} {\bibfnamefont {A.~J.}\ \bibnamefont {Tolley}},\
  }\href {\doibase 10.1103/PhysRevLett.106.231101} {\bibfield  {journal}
  {\bibinfo  {journal} {Phys. Rev. Lett.}\ }\textbf {\bibinfo {volume} {106}},\
  \bibinfo {pages} {231101} (\bibinfo {year} {2011})},\ \Eprint
  {http://arxiv.org/abs/1011.1232} {arXiv:1011.1232 [hep-th]} \BibitemShut
  {NoStop}%
\bibitem [{\citenamefont {Coleman}\ \emph {et~al.}(1992)\citenamefont
  {Coleman}, \citenamefont {Preskill},\ and\ \citenamefont
  {Wilczek}}]{Coleman:1991ku}%
  \BibitemOpen
  \bibfield  {author} {\bibinfo {author} {\bibfnamefont {S.~R.}\ \bibnamefont
  {Coleman}}, \bibinfo {author} {\bibfnamefont {J.}~\bibnamefont {Preskill}}, \
  and\ \bibinfo {author} {\bibfnamefont {F.}~\bibnamefont {Wilczek}},\ }\href
  {\doibase 10.1016/0550-3213(92)90008-Y} {\bibfield  {journal} {\bibinfo
  {journal} {Nucl. Phys.}\ }\textbf {\bibinfo {volume} {B378}},\ \bibinfo
  {pages} {175} (\bibinfo {year} {1992})},\ \Eprint
  {http://arxiv.org/abs/hep-th/9201059} {arXiv:hep-th/9201059 [hep-th]}
  \BibitemShut {NoStop}%
\bibitem [{\citenamefont {Kimura}\ \emph {et~al.}(2012)\citenamefont {Kimura},
  \citenamefont {Kobayashi},\ and\ \citenamefont {Yamamoto}}]{Kimura:2011dc}%
  \BibitemOpen
  \bibfield  {author} {\bibinfo {author} {\bibfnamefont {R.}~\bibnamefont
  {Kimura}}, \bibinfo {author} {\bibfnamefont {T.}~\bibnamefont {Kobayashi}}, \
  and\ \bibinfo {author} {\bibfnamefont {K.}~\bibnamefont {Yamamoto}},\ }\href
  {\doibase 10.1103/PhysRevD.85.024023} {\bibfield  {journal} {\bibinfo
  {journal} {Phys. Rev.}\ }\textbf {\bibinfo {volume} {D85}},\ \bibinfo {pages}
  {024023} (\bibinfo {year} {2012})},\ \Eprint {http://arxiv.org/abs/1111.6749}
  {arXiv:1111.6749 [astro-ph.CO]} \BibitemShut {NoStop}%
\bibitem [{\citenamefont {Babichev}\ \emph {et~al.}(2015)\citenamefont
  {Babichev}, \citenamefont {Charmousis},\ and\ \citenamefont
  {Hassaine}}]{Babichev:2015rva}%
  \BibitemOpen
  \bibfield  {author} {\bibinfo {author} {\bibfnamefont {E.}~\bibnamefont
  {Babichev}}, \bibinfo {author} {\bibfnamefont {C.}~\bibnamefont
  {Charmousis}}, \ and\ \bibinfo {author} {\bibfnamefont {M.}~\bibnamefont
  {Hassaine}},\ }\href {\doibase 10.1088/1475-7516/2015/05/031} {\bibfield
  {journal} {\bibinfo  {journal} {JCAP}\ }\textbf {\bibinfo {volume} {1505}},\
  \bibinfo {pages} {031} (\bibinfo {year} {2015})},\ \Eprint
  {http://arxiv.org/abs/1503.02545} {arXiv:1503.02545 [gr-qc]} \BibitemShut
  {NoStop}%
\bibitem [{\citenamefont {Goldberger}\ and\ \citenamefont
  {Rothstein}(2006)}]{Goldberger:2004jt}%
  \BibitemOpen
  \bibfield  {author} {\bibinfo {author} {\bibfnamefont {W.~D.}\ \bibnamefont
  {Goldberger}}\ and\ \bibinfo {author} {\bibfnamefont {I.~Z.}\ \bibnamefont
  {Rothstein}},\ }\href {\doibase 10.1103/PhysRevD.73.104029} {\bibfield
  {journal} {\bibinfo  {journal} {Phys. Rev.}\ }\textbf {\bibinfo {volume}
  {D73}},\ \bibinfo {pages} {104029} (\bibinfo {year} {2006})},\ \Eprint
  {http://arxiv.org/abs/hep-th/0409156} {arXiv:hep-th/0409156 [hep-th]}
  \BibitemShut {NoStop}%
\bibitem [{\citenamefont {Goldberger}(2007)}]{Goldberger:2007hy}%
  \BibitemOpen
  \bibfield  {author} {\bibinfo {author} {\bibfnamefont {W.~D.}\ \bibnamefont
  {Goldberger}},\ }in\ \href@noop {} {\emph {\bibinfo {booktitle} {{Les Houches
  Summer School - Session 86: Particle Physics and Cosmology: The Fabric of
  Spacetime Les Houches, France, July 31-August 25, 2006}}}}\ (\bibinfo {year}
  {2007})\ \Eprint {http://arxiv.org/abs/hep-ph/0701129} {arXiv:hep-ph/0701129
  [hep-ph]} \BibitemShut {NoStop}%
\bibitem [{\citenamefont {Hiramatsu}\ \emph {et~al.}(2013)\citenamefont
  {Hiramatsu}, \citenamefont {Hu}, \citenamefont {Koyama},\ and\ \citenamefont
  {Schmidt}}]{Hiramatsu:2012xj}%
  \BibitemOpen
  \bibfield  {author} {\bibinfo {author} {\bibfnamefont {T.}~\bibnamefont
  {Hiramatsu}}, \bibinfo {author} {\bibfnamefont {W.}~\bibnamefont {Hu}},
  \bibinfo {author} {\bibfnamefont {K.}~\bibnamefont {Koyama}}, \ and\ \bibinfo
  {author} {\bibfnamefont {F.}~\bibnamefont {Schmidt}},\ }\href {\doibase
  10.1103/PhysRevD.87.063525} {\bibfield  {journal} {\bibinfo  {journal} {Phys.
  Rev.}\ }\textbf {\bibinfo {volume} {D87}},\ \bibinfo {pages} {063525}
  (\bibinfo {year} {2013})},\ \Eprint {http://arxiv.org/abs/1209.3364}
  {arXiv:1209.3364 [hep-th]} \BibitemShut {NoStop}%
\bibitem [{\citenamefont {Dvali}\ \emph {et~al.}(2003)\citenamefont {Dvali},
  \citenamefont {Gruzinov},\ and\ \citenamefont {Zaldarriaga}}]{Dvali:2002vf}%
  \BibitemOpen
  \bibfield  {author} {\bibinfo {author} {\bibfnamefont {G.}~\bibnamefont
  {Dvali}}, \bibinfo {author} {\bibfnamefont {A.}~\bibnamefont {Gruzinov}}, \
  and\ \bibinfo {author} {\bibfnamefont {M.}~\bibnamefont {Zaldarriaga}},\
  }\href {\doibase 10.1103/PhysRevD.68.024012} {\bibfield  {journal} {\bibinfo
  {journal} {Phys. Rev.}\ }\textbf {\bibinfo {volume} {D68}},\ \bibinfo {pages}
  {024012} (\bibinfo {year} {2003})},\ \Eprint
  {http://arxiv.org/abs/hep-ph/0212069} {arXiv:hep-ph/0212069 [hep-ph]}
  \BibitemShut {NoStop}%
\end{thebibliography}%

\end{document}